%% file: main.tex
\documentclass[aps,prx,twocolumn,superscriptaddress,showkeys,nofootinbib,floatfix]{revtex4-2}

\usepackage{graphicx}
\usepackage[export]{adjustbox}
\usepackage{color,colortbl}
\definecolor{Gray}{gray}{0.9}
\usepackage{multirow}
\usepackage{comment}

\def\arraystretch{1.5}

\usepackage{subcaption}
\usepackage{siunitx}                   % Proper formatting for SI units
\usepackage[version=4]{mhchem}	       % Chemical formulae

\usepackage{hyperref}
\usepackage{braket}
\usepackage{physics}

\usepackage{xcolor}

\begin{document}

%\preprint{}

\title{Sensor-assisted fault mitigation in quantum computation}
%\thanks{Work supported by no one, in fact.}


%\author{}
%\thanks{}
%\affiliation{}

\author{John L.\ Orrell} \email[Corresponding author: ]{john.orrell@pnnl.gov}
\affiliation{Pacific Northwest National Laboratory, Richland, WA 99352, USA}
% https://orcid.org/0000-0001-7968-4051

\author{Ben Loer}
\affiliation{Pacific Northwest National Laboratory, Richland, WA 99352, USA}
% https://orcid.org/0000-0002-7096-5911


\date{\today}

\begin{abstract}
We propose a method to assist fault mitigation in quantum computation through the use of sensors co-located near physical qubits. Specifically, we consider using transition edge sensors co-located on silicon substrates hosting superconducting qubits to monitor for energy injection from ionizing radiation, which has been demonstrated to increase decoherence in transmon qubits. We generalize from these two physical device concepts and explore the potential advantages of co-located sensors to assist fault mitigation in quantum computation. In the simplest scheme, co-located sensors beneficially assist rejection of calculations potentially affected by environmental disturbances. Investigating the potential computational advantage further required development of an extension to the standard formulation of quantum error correction. In a specific case of the standard three-qubit, bit-flip quantum error correction code, we show that given a 20\% overall error probability per qubit, approximately 90\% of repeated calculation attempts are correctable. However, when \emph{sensor-detectable} errors account for 45\% of overall error probability, the use of co-located sensors uniquely associated with independent qubits boosts the fraction of correct final-state calculations to 96\%, at the cost of rejecting 7\% of repeated calculation attempts.

%\begin{description}
%\item[DOI] \href{https://doi.org/10.1103/PRXQuantum.x.yyyyyy}{10.1103/PRXQuantum.x.yyyyyy}
%\item[Subject Areas] Interdisciplinary Physics, Quantum Information
%\end{description}
\end{abstract}

%\keywords{Superconducting qubits; Transition edge sensors; Quantum fault tolerance; Quantum error correction}

\maketitle

%\section{}
%\subsection{}
%\subsubsection{}

%%
%% INTRODUCTION
%%

\section{\label{sec:intro}Introduction}

Many mechanisms may lead to state decoherence in the physical implementation of quantum computing systems. Recent reports \cite{PhysRevLett.121.117001,Oliver2020,Cardani2020} show deleterious effects in superconducting kinetic inductance devices and superconducting transmon qubits correlated with ionizing radiation levels, identifying yet another mechanism causing decoherence. As others have~\cite{Cardani2019}, we postulate these observed phenomena stem from the same underlying process: the instantaneous injection of energy into the superconducting device and the device's substrate as a result of impinging ionizing radiation. It is possible to reduce the rate of ionizing radiation energy injections by shielding against naturally occurring radiation sources in the laboratory and by placing systems underground to shield against cosmic rays. These techniques are commonly employed for rare event searches in nuclear and particle physics research, including searches operating at mK temperatures~\cite{PhysRevD.95.082002,Armengaud2017,PhysRevD.100.102002,ISI:000386879300001,ALDUINO20199,ISI:000475616600001}. However, the history of such physics research experiments demonstrate it is difficult to entirely shield against the ionizing radiation present in any instrumentation laboratory. Thus, we contemplate superconducting qubit operation in a regime of low, but non-zero, rates of ionizing-radiation-induced energy injections. From there we draw an inference to a superconducting qubit device concept employing the use of co-located sensors that can signal when an ionizing radiation energy injection has occurred, signifying probable error in the quantum computation.

We employ the terminology \emph{fault mitigation in quantum computation} to distinguish from purely \emph{quantum} computational means for achieving fault tolerance or error correction~\cite{7167244}. In the simplest application of our device concepts, we show co-located sensors can provide modest fault mitigation through selective result-acceptance in redundant (``many shot'') computation schemes, where the same quantum calculation is repeated multiple times. Speculatively, as this will require advances in superconducting qubit interconnection techniques, we explore how co-located sensors can identify uncorrectable errors within the framework of quantum error correction codes.

%%
%% TES-ASSISTED QEC DEVICE CONCEPT
%%

\section{\label{sec:TES-assisted_qubit_concept}TES-assisted qubit device concept}

\begin{figure*}[htb!]
  %\centering
  \includegraphics[width=\textwidth]{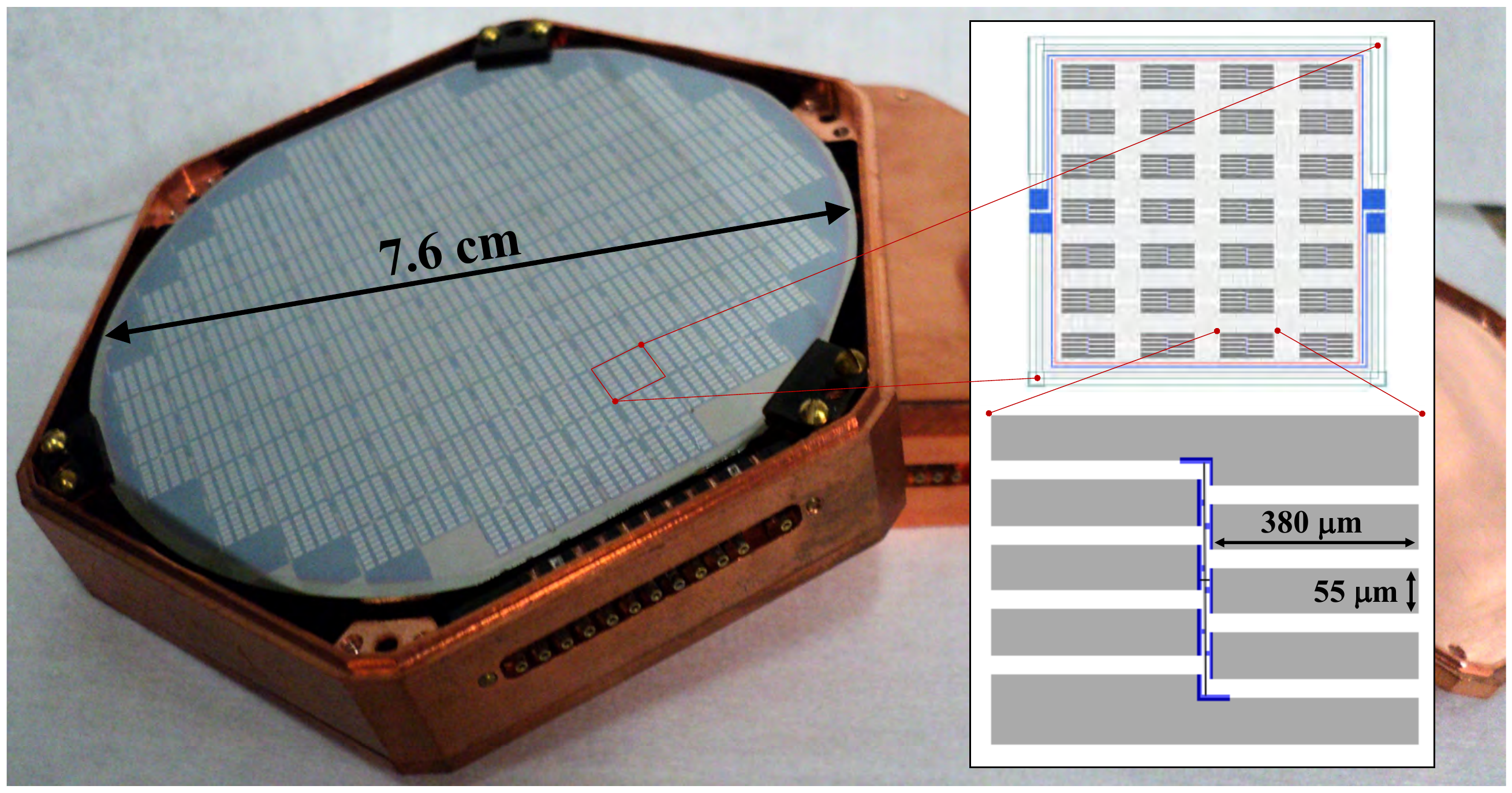}
  %%% BIG/SMALL
\caption{Photograph of a CDMS II ZIP detector contained within it's hexagonal copper housing~\cite{CDMS-iZIP-photo}. A rectangular patch of Quasiparticle-trap-assisted Electrothermal-feedback Transition-edge-sensors (QETs) is highlighted on the detector's surface and shown schematically in more detail to the right. The upper right schematic shows a $7\times4$ array of individual QETs. To the lower right, a single QET is shown with the physical dimensions of the aluminum phonon/quasiparticle collector fins. The TES consists of the smaller blue portions of the circuit, located in the center, aligned vertically and touching each of the six collector fins as well as the voltage rails, located top and bottom.}
\label{fig:cdms-ii-zip}
\end{figure*}

This section presents a notional concept for the physical implementation of devices combining ionizing radiation transition edge sensors (TES) and superconducting qubits that share a common silicon substrate.

% TES
\subsection{\label{sec:TES}Transition edge sensor devices}

In a TES~\cite{ISI:000231009400003}, the material's effective temperature is set such that the material resides on the ``transition edge'' between the superconducting and normal conducting states. Any additional energy added to the material will increase the temperature and push the TES toward the normal conducting phase, dramatically raising the electrical resistance of the material. Sensing this change in resistance in a circuit makes the TES useful for detecting small amounts of absorbed energy.

A key step in the development~\cite{doi:10.1063/1.1770037,Irwin2005} of TES devices as practical sensors was the use of direct current (DC) voltage bias to provide negative electrothermal feedback (ETF) to stabilize the readout circuit~\cite{doi:10.1063/1.113674}. As diagrammed in the cited seminal reference, superconducting quantum interference devices (SQUIDs) are typically used to monitor the resistance-dependent current in the ETF TES circuit through a current-induced magnetic field. While the TES may reside at tens of mK~temperatures in the ``mixing chamber'' stage of a refrigerator, the SQUIDs monitoring the circuit are typically located at a warmer stage, often at a $\simeq600$~mK ``still'' stage~\cite{AKERIB2008476}. This provides for physical separation and magnetic shielding between the TES devices and the SQUIDs.

\begin{figure*}[hbt!]
\begin{subfigure}{.314\textwidth}
  \centering
  \includegraphics[width=\textwidth]{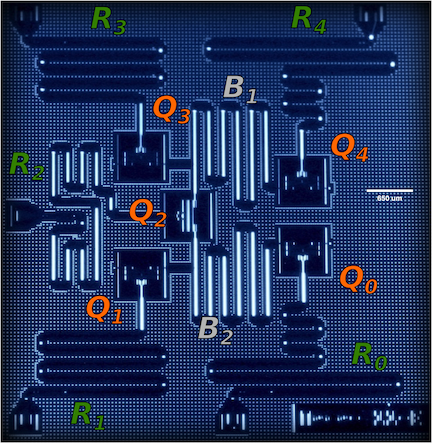}
  %%% BIG/SMALL
  \caption{Micrograph of IBM 5-qubit device.}
  \label{fig:ibmqx2_yorktown_microgrpah}
\end{subfigure}
\begin{subfigure}{.32\textwidth}
  \centering
  \includegraphics[width=\textwidth]{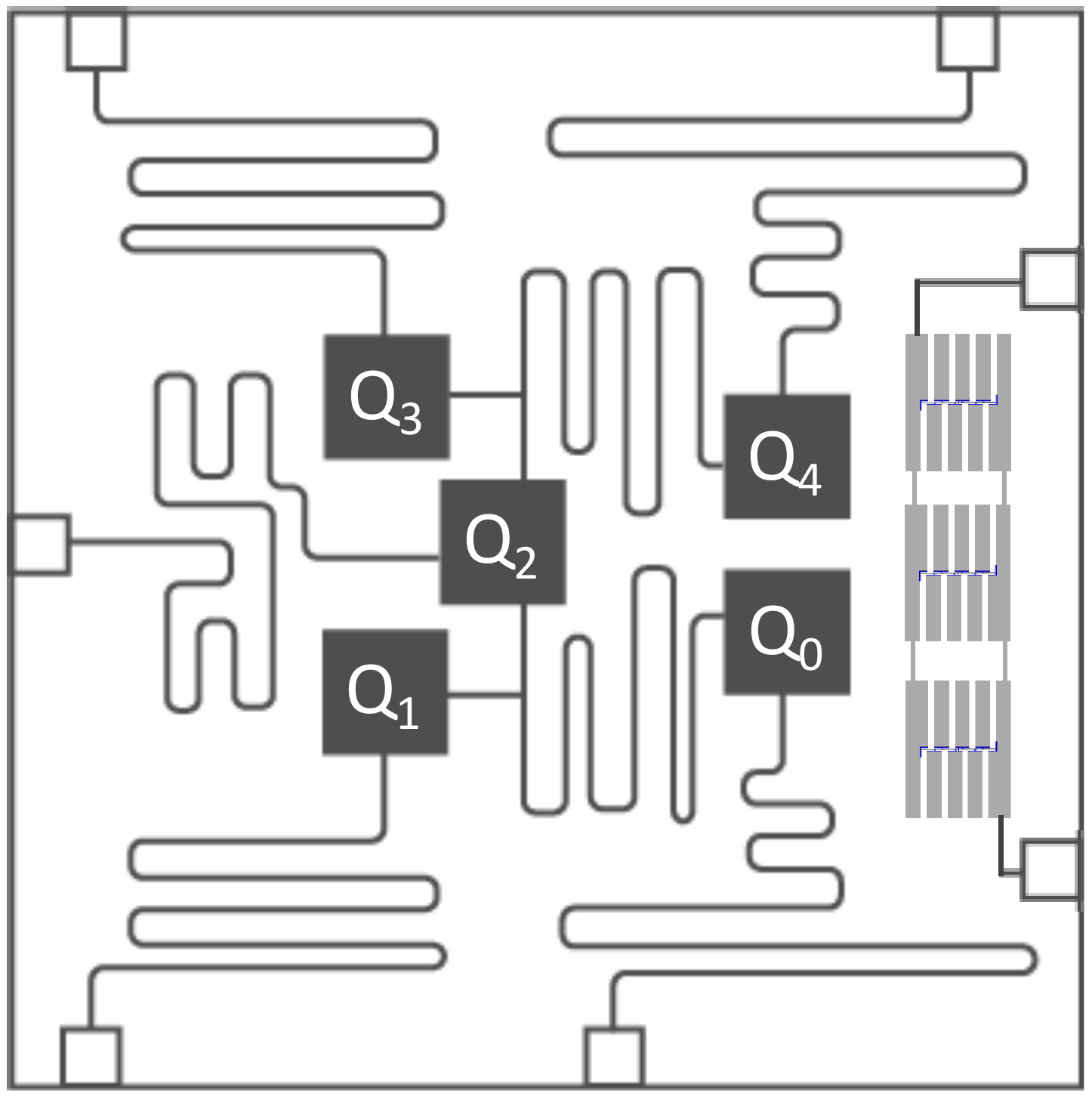}
  \caption{IBM 5-qubit scheme with 3 QETs.}
  \label{fig:ibmqx2_yorktown_schematic}
\end{subfigure}
\begin{subfigure}{.32\textwidth}
  \centering
  \includegraphics[width=\textwidth]{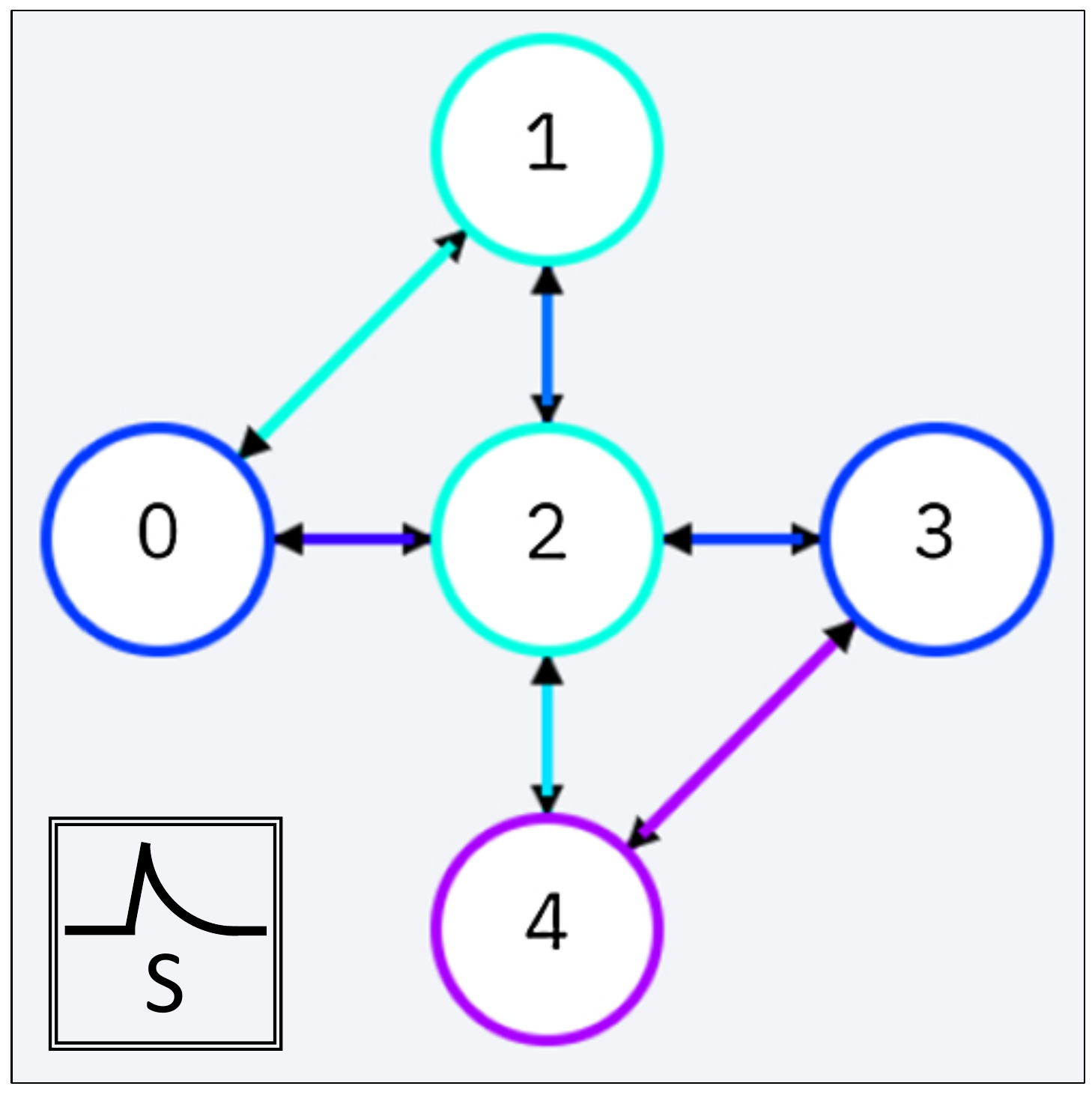}
  \caption{Gate connections with sensor patch.}
  \label{fig:ibmqx2_yorktown_connections}
\end{subfigure}
\caption{Labeled micrograph of the IBM 5-qubit \texttt{ibmqx2} Yorktown backend device (\ref{fig:ibmqx2_yorktown_microgrpah})~\cite{ibmqx2_yorktown}. The white, horizontal bar located to the center-right within the micrograph provides a length scale of 650~$\mu$m. Resonators ($R_{\#}$), qubits ($Q_{\#}$), and bus resonators ($B_{\#}$) are labeled on the micrograph image. A schematic of the Yorktown backend is shown in (\ref{fig:ibmqx2_yorktown_schematic}). On the right-hand edge of the 5-qubit schematic, we have overlaid three QET devices (see Fig.~\ref{fig:cdms-ii-zip}), which can serve as a co-located ionizing radiation sensor. The connection map for the Yorktown backend is shown in (\ref{fig:ibmqx2_yorktown_connections}). We have added a representation of a co-located sensor, labeled as ``S'' with a symbolic event pulse contained within a double-lined box suggesting the classical nature of the sensor device.}
\label{fig:ibmqx2_yorktown}
\end{figure*}

TES sensors developed by the SuperCDMS collaboration~\cite{PhysRevD.95.082002} employ QET devices, defined as Quasiparticle-trap-assisted Electrothermal feedback Transition-edge sensor devices~\cite{doi:10.1063/1.1146105}. In these QET devices, superconducting aluminum films are deposited on Ge or Si crystals, in contact with the tungsten-based ETF TES devices. Phonon energy present in the crystal substrate breaks Cooper pairs in the superconducting Al films. The resultant quasiparticles diffuse through the Al film to the W-based ETF TES, ultimately resulting in a TES transition event used for event detection. Typically, multiple QET devices are operated in parallel in a circuit to provide increased phonon energy collection coverage with a single sensor channel.

Figure~\ref{fig:cdms-ii-zip} shows a CDMS ZIP (Z-dependent Ionization- and Phonon-mediated) detector~\cite{CDMS-iZIP-photo}. We added the schematics~\cite{PhysRevD.72.052009}, scale overlays, and highlighting lines. Detailed descriptions of lithographic fabrication techniques for similar devices are available~\cite{JASTRAM201514}. It is worth noting the QET devices used in these detector applications are essentially ``classical'' signal sensors. That is, the TES circuit operates through a process of Joule heating of a material in response to a thermalizing population of quasiparticles, produced by a population of thermal and athermal substrate phonons.

% Qubits
\subsection{\label{sec:superconducting_qubit}Superconducting qubit devices}

There are many modalities for the physical implementation of qubits.
These modalities include trapped ions, superconducting circuits, photon systems manipulated either with linear optics or quantum dots, neutral atoms, semiconductor devices typified by either optically active nitrogen vacancies in diamond or electronically manipulated electron spins, and most recently topological qubits that are based in collective properties of solid state systems~\cite{NAP25196}. In all cases, the goal is to isolate a physical two-level quantum system that can be manipulated for quantum computation. In this report we focus on superconducting qubit devices.

In this work we consider transmon qubits~\cite{PhysRevA.76.042319} based on our experience with them in studies of the effect of ionizing radiation on their coherence time~\cite{Oliver2020}. Furthermore, the IBM Q Experience~\cite{IBMQ} provides access to transmon-based multi-qubit devices~\cite{ISI:000399429500002,ISI:000542630400002} for cloud-based quantum computing. We use these resources as a reference for exploring sensor-assisted fault mitigation in quantum computation.

Figure~\ref{fig:ibmqx2_yorktown_schematic} shows in schematic representation the combining of three QET devices with the qubit chip layout. The schematic diagram (Fig.~\ref{fig:ibmqx2_yorktown_schematic}) captures our proposed hybrid sensor and qubit device concept. A notional connectivity diagram (Fig.~\ref{fig:ibmqx2_yorktown_connections}) further abstracts the generalized idea of a co-located sensor for detection of environmental disturbances.

% TES-assisted devices
\subsection{\label{sec:TES-assisted_qubit_devices}TES-assisted qubit devices}

A hybrid device as suggested by Figure~\ref{fig:ibmqx2_yorktown_schematic} is producible with today's fabrication techniques. Furthermore, we do not foresee any inherent incompatibility in co-operation of the DC voltage biased QET devices and the microwave frequency controls of the qubits. Specifically, QET devices on a silicon chip are operated using a DC voltage bias across the TES of approximately 40~mV. From the TES-SQUID circuit's quiescent state, ionizing radiation induced events appear as $\simeq$5~$\mu$s rising-edge current excursions of $\simeq$100~nA amplitudes and $\simeq$100~$\mu$s pulse decay times. These representative operational details are derived from the SuperCDMS HVeV chip-scale dark matter search devices~\cite{PhysRevLett.121.051301,doi:10.1063/1.5010699}.

The above described QET operating characteristics are in contrast to transmon qubit operation following the theory of circuit quantum electrodynamics (cQED)~\cite{Schuster2007}. Qubits are typically controlled via radiofrequency pulses applied through co-planar waveguide microwave transmission lines, typically in the $\simeq$5~GHz range. Specifically, qubits are coupled to the transmission line via superconducting Al circuit meander resonators designed to have unique resonance frequencies in the same $\simeq$5~GHz range, resulting from the details of their physical shape. Each qubit's resonance frequency is designed to lie off-resonance (detuned) from the paired resonator's resonance frequency to allow dispersive readout from the qubit via the resonator~\cite{doi:10.1063/1.5089550}. Multiple such qubit-resonator pairs can exist on the same silicon chip and even connect to same transmission line~\cite{Jerger_2011}, so long as all resonance frequencies are fully offset. The $\simeq$5~GHz RF control pulses are typically $\simeq$10s of nanoseconds duration and have millivolt scale amplitudes at the readout resonator, resulting in $\simeq$100s of nanoamperes of current in the qubit circuit.

The hybrid devices we envision, having the above described characteristics, would consist of QET and transmon qubit devices simultaneously operated at roughly 30--50~mK. There are two obvious possible ``cross-talk'' scenarios between the QETs and the quibts. The first is through near-resonance coupling of RF qubit control pulses in the QET. We believe the QET physical layout can be optimized to reduce the potential for this coupling. It is not obvious current excursions in the QET devices would have any coupling to the qubit circuits. The second ``cross-talk'' mechanism is through quasiparticle generation via power input from either device type. There is ample evidence from the operation of arrays of QET and superconducting qubit devices that each device type can be operated without substantial injection of thermal energy into the substrate, which would result in elevated quasiparticle levels in the superconducting circuits of either device type. We are not aware of any conceptually similar device created to date to that experimentally tests the veracity of these claims.

In the next section, we assess the potential value of such a co-located sensor in contributing to fault mitigation in quantum computations. The initial evaluation considers plausible devices we believe can be fabricated today. Such devices would likely employ co-located sensors in a ``veto'' role to reject computations suspected of excessive error-inducing environmental disturbances. Taking the assessment a step further, we speculate on the error correction performance of independent qubits systems, where each qubit is uniquely associated with an individual co-located sensor. In the case of superconducting qubits, this idealization would manifest in the case where QET-qubit pairs each reside on separate silicon substrate chips and are potentially interconnected through superconducting air-bridges or capacitive coupling across gaps between chips. We note the choice of the class of TES/QET devices~\cite{Ullom_2015} for the co-located sensor is potentially interchangeable with microwave kinetic inductance detectors (MKIDs)~\cite{Day_2003} or superconducting nanowire detectors~\cite{Natarajan_2012}.

%%
%% ERROR CORRECTION
%%

\section{\label{sec:error_estimation}Quantum error mitigation}

Pedagogical development of qubit-based quantum error correction considers two complementary forms of error: bit-flip error and sign-flip error. Within the Bloch sphere picture of a qubit, these errors correspond to state error and phase error. These two flip-type quantum errors are highly idealized \emph{binary symmetric channel} representations of the otherwise continuous error experienced by real qubits~\cite{Devitt_2013}. We note ionizing radiation induced error in superconducting transmon qubits is almost certainly a continuous noise source best represented by arbitrary three-angle unitary transformations (or much worse). However, for our goal of developing an intuition for the relative utility of sensor-assisted error mitigation in quantum computation, we will focus solely on bit-flip errors, to the exclusion of all others. This assumption and other assumptions we make in the following developments are assessed in the Discussion section.

Our goal is to determine how information gained from a co-located sensor---\emph{without performing any measurement on the quantum computation qubit(s)}---can assist in the implementation of error mitigation in quantum computation. We begin with the hybrid device concept presented in Section~\ref{sec:TES-assisted_qubit_concept}, Figure~\ref{fig:ibmqx2_yorktown_schematic}. For illustrative purposes, we make use of the IBM Quantum Experience~\cite{ibmqx2_yorktown} as a source of some realistic scenarios, specifically working with the Yorktown (\texttt{ibmqx2}) 5-qubit backend~\cite{PhysRevLett.109.240504}. We will refer to this simply as the ``Yorktown backend'' for brevity. We conclude by investigating a fully abstracted hypothetical case when co-located sensors are uniquely assigned to individual, independent qubits.

\begin{figure}[t!]
\includegraphics[width=\columnwidth]{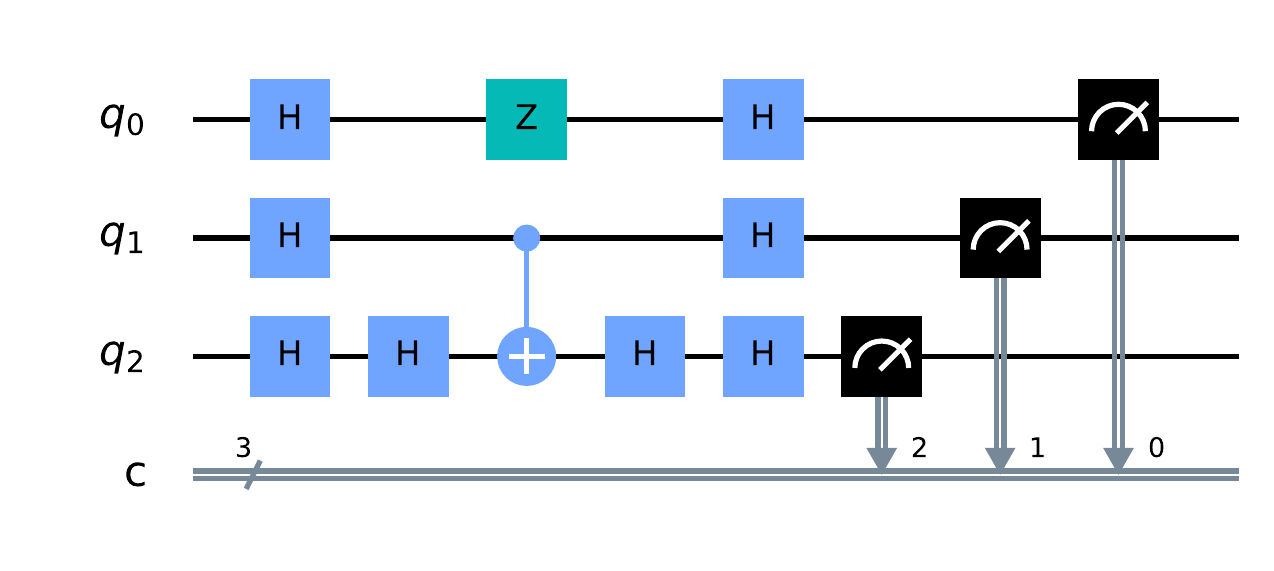}
\caption{A simple balanced Deutsch-Jozsa calculation used as a test case for investigating the role of co-located sensors in calculations performed by devices such as the IBM 5-qubit \texttt{ibmqx2} Yorktown backend (see Fig.~\ref{fig:ibmqx2_yorktown}).}
\label{fig:balanced-dj-circuit}
\end{figure}

\begin{figure}[htb!]
\raggedright
\begin{center}
    \underline{\textbf{Balanced Deutsch-Jozsa calculation results}} \\
\end{center}
(a)~\vspace{-1.6em} \\
\begin{subfigure}{\columnwidth}
  \includegraphics[width=\columnwidth,center]{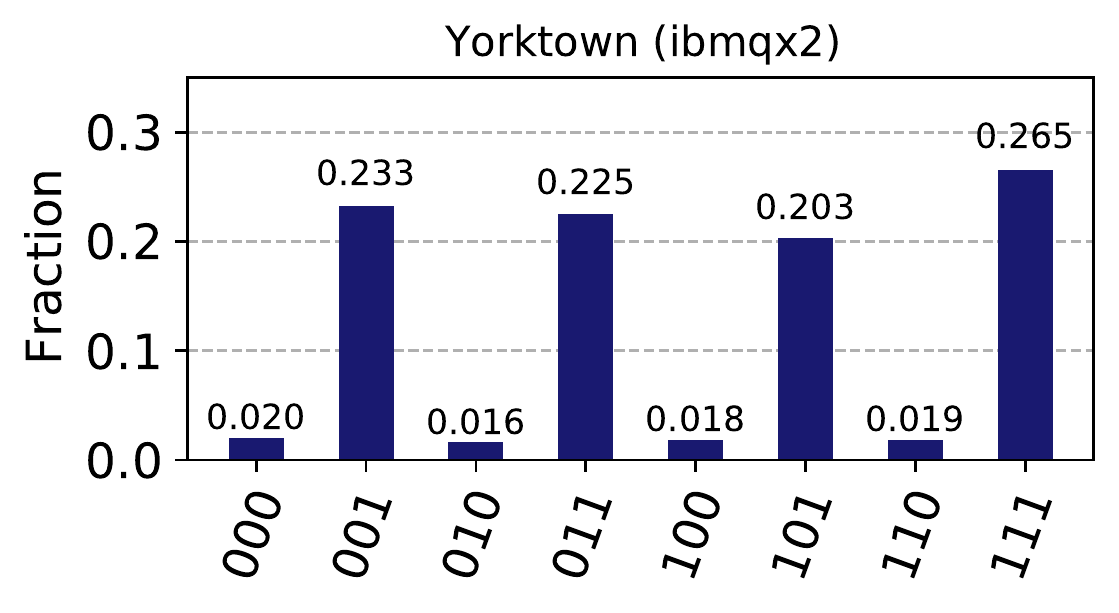}
\end{subfigure}
(b)~\vspace{-1.6em} \\
\begin{subfigure}{\columnwidth}
  \includegraphics[width=\columnwidth,center]{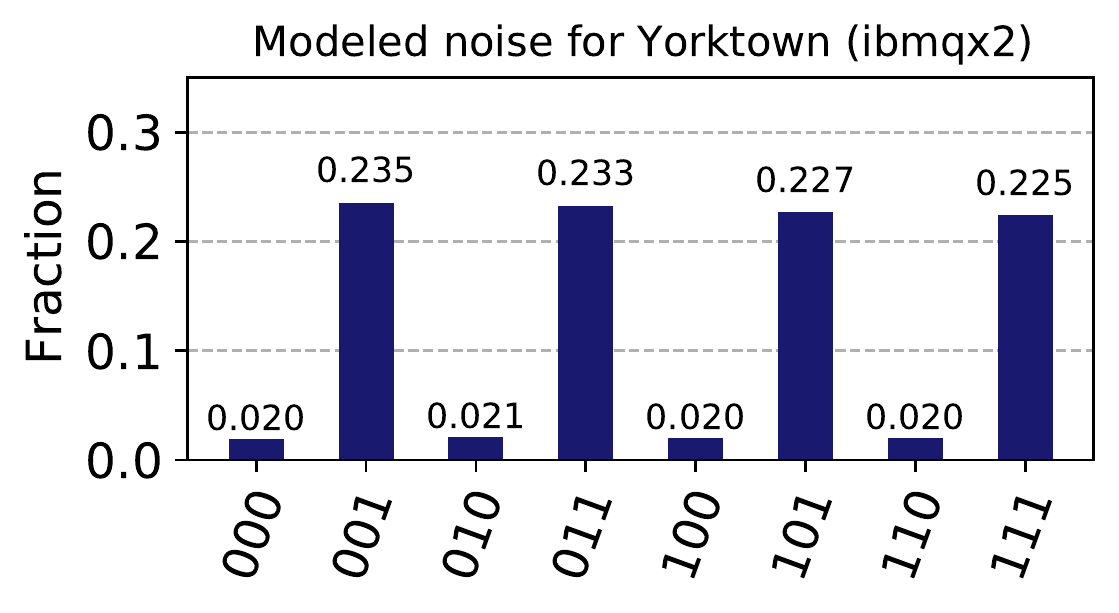}
\end{subfigure}
(c)~\vspace{-1.6em} \\
\begin{subfigure}{\columnwidth}
  \includegraphics[width=\columnwidth,center]{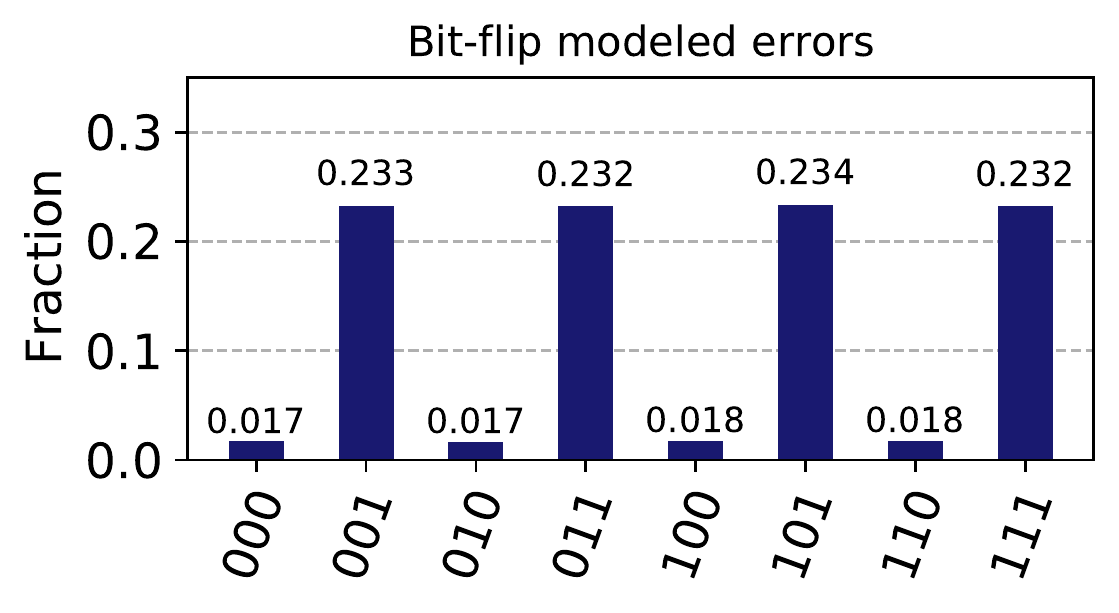}
\end{subfigure}
(d)~\vspace{-1.6em} \\
\begin{subfigure}{\columnwidth}
  \includegraphics[width=\columnwidth,center]{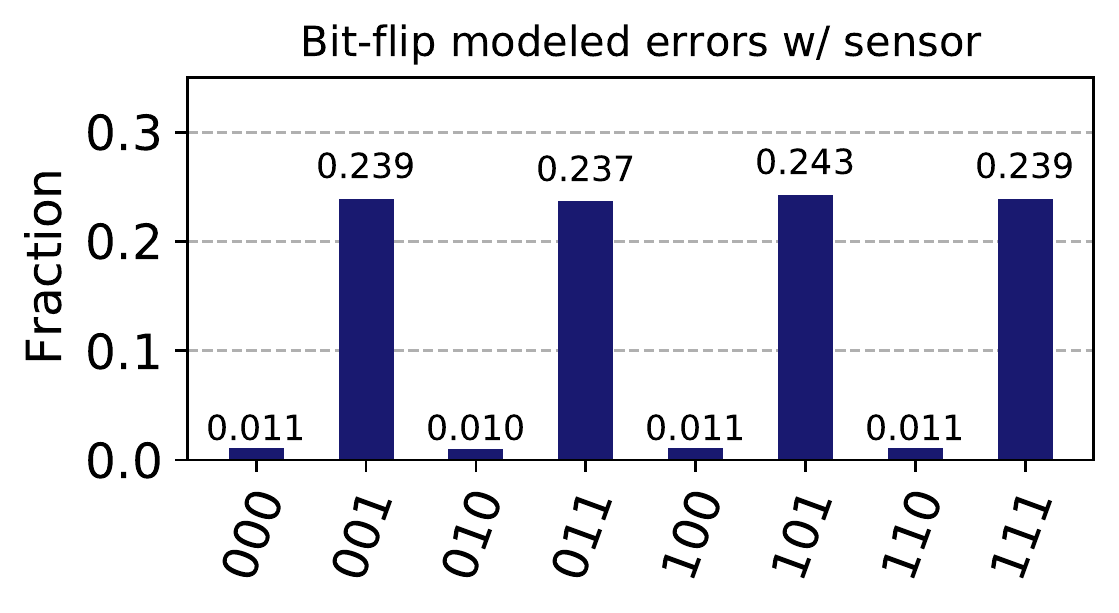}
\end{subfigure}\
\caption{Results from three implementations of a balanced Deutsch-Jozsa calculation (See Fig.~\ref{fig:balanced-dj-circuit}). Results (c) and (d) are the same calculation without and with an assumed co-located sensor for environmental error-inducing event detection, respectively. Detection of 40\% of single bit-flip errors with a classical co-sensor reduces the number of faulty final states}
\label{fig:balanced-dj}
\end{figure}.

% Example - Yorktown
\subsection{\label{sec:example_error}Example calculation: Repetition and error}

Quantum error correction is often presented as an approach toward the correction of errors in an idealized, \emph{single-pass} quantum computation calculation. The application of quantum computation routinely uses computational repetition (repeating the same calculation many times) to achieve averaged results that approach the idealized, single-pass calculation result for large numbers of repetitions. Furthermore, a single-pass quantum calculation is only able to return the ``correct'' answer in cases where the result is uniquely identifiable with a single eigenvector of the measurement basis. More generally, in cases analogous to quantum phase estimation and/or quantum state tomography, the relative weight of the measurement basis eigenvectors---determined through computational repetition---is key to determining the underlying quantum state. Thus, in quantum calculations, computational repetition is used advantageously in \emph{both} statistical averaging for error mitigation \emph{and} quantum state estimation as part of the underlying calculation method. In addition, and in entire generality, if erroneous final states are identifiable within this repetition process, then either better accuracy is obtained for a fixed number of repetitions or the same accuracy is achievable with fewer repetitions.

Figure~\ref{fig:balanced-dj}(a) shows the results from 81,920~repetitions of a simple balanced Deutsch-Jozsa calculation (see Fig.~\ref{fig:balanced-dj-circuit}) implemented on the Yorktown backend. The ``correct'' result is equal weight in each of the four states $|001|$, $|011|$, $|101|$, and $|111|$ (i.e., 25\% of the 81,920~trials in each of the four states), with statistical fluctuations from the finite sample size. However, the data report \emph{at minimum} 5,734 trials of the repeated quantum calculation were in error, reporting measurements of the states $|000|$, $|010|$, $|100|$, and $|110|$.

We contemplate the possibility that \emph{some} of the error states are the result of ionizing radiation striking the Yorktown backend during the computational repetitions. Our prior work~\cite{Oliver2020} suggests the actual fraction of ionizing radiation disturbances is small for devices such as the Yorktown backend. However, for the sake of intellectual exploration, we wish to consider when some significant percentage of the induced error states are due to ionizing radiation or some other environmental disturbance detectable by a co-located sensor. We are thus implicitly assuming some error inducing phenomenon are also \emph{not} detectable by the co-located sensor, as is normally assumed in quantum error correction schemes. For concreteness, we consider a case where 60\% of the errors are \underline{\emph{not}} due to ionizing radiation (or some other environmental disturbance), which is detectable by a co-located sensor on the qubit chip.

We have no method for assessing the true error cases for any particular computational repetition of the Yorktown backend, so we must create a model of the noise. The Qiskit programming language provides a mechanism for simulating the noise of a specific backend device, based on measured gate error rates and coherence times. Fig~\ref{fig:balanced-dj} shows the results of many such calculations performed during the week of 12 October 2020. Unfortunately, we are not aware of a way to use the Qiskit modeled noise to determine for a single repetition of the calculation when an error may have been induced (modeled) for a qubit. Thus, we created a simple bit-flip-based noise model simulation designed to \emph{mimic} the statistical properties of the Yorktown backend performing the balanced Deutsch-Jozsa calculation.

We assign a single bit-flip (\textbf{\texttt{X}}-gate) to follow each of the eleven operations on the qubits in the circuit diagram of Figure~\ref{fig:balanced-dj-circuit}, including the control qubit on \textit{qubit}$_1$. We find setting the bit-flip error probability to 7\% in this highly over-simplified model simulation roughly reproduces the balanced Deutsch-Jozsa calculation's statistical distribution of results seen on the actual Yorktown backend device. Thus, we now have a method for determining within a single repetition of the calculation when an error was induced within the quantum circuit by any one (or more) of the bit-flip errors. We simulated 81,920 single-shot calculations, where each time a balanced Deutsch-Jozsa circuit was created with a randomly generated set of bit-flip errors contained within the circuit, based upon the 7\% gate error probability mentioned above. The results are shown in Fig~\ref{fig:balanced-dj}(c). Recall, rather than assuming 100\% of the induced errors are due to an environmental disturbance that can be detected by the co-located sensor, we instead assume 60\% of the errors are \underline{\emph{not}} detectable by the co-located sensor. 

Fig~\ref{fig:balanced-dj}(d) shows the results when the co-located sensor would provide information to reject a number of the calculations (20,282 shots in this case) that are expected to potentially be in error. This improves the performance of the quantum calculation, showing a reduction of the fraction of calculation repetitions reporting states $|000|$, $|010|$, $|100|$, and $|110|$ compared to that shown in Fig~\ref{fig:balanced-dj}(c). Our first substantial conclusion is that this improvement is at the cost of rejecting outright a number of the calculations from consideration. We repeat, the calculation improves because those calculations with the potential for being environmentally disturbed are preferentially rejected from consideration in calculating the final results after all repetitions are complete. The Appendix to this report further investigates the statistical properties of the results shown in Fig~\ref{fig:balanced-dj}.

The form of error mitigation described above is of the simplest variety. The co-located sensor provides a case-by-case capacity to reject or ``veto'' individual, ``single-shot'' calculations. At the expense of throwing-away the so-flagged calculation trials, it is possible to improve the numerical accuracy of quantum calculations employing repetition for purposes of result averaging or quantum state determination via the measurement eigenvector weightings. While these improvements are modest, we believe devices such as that described by Figure~\ref{fig:ibmqx2_yorktown_schematic} can be fabricated today and take advantage of sensors to selectively reject calculations where environmental phenomenon have potentially disturbed the quantum computational system.

% Error types
\subsection{\label{sec:two_error_types}Error types: Environmental and entangling}

We now propose to distinguish more clearly between two classes of phenomenon resulting in quantum decoherence of qubit systems. In this discussion, we have in mind superconducting qubit devices, but we believe these definitions are sufficiently general as to apply to other physical implementations of qubits. We suggest framing two types of qubit error generation mechanisms that can appear in physical qubit systems: (1) Environmental disturbances and (2) effects having quantum entanglement. These two types are not mutually exclusive, but they should be exhaustive. As such, we warily adopt substantively \emph{different} meaning for the terms ``environment'' and ``environmental,'' due to a lack of better terminology. We acknowledge our use of these terms may seem counter to the sense used by other authors. 

For this report we consider environmental error-inducing disturbances as those phenomena that are \emph{independent} of the presence or absence of a qubit state. In a superconducting qubit device, we have in mind phenomena such as energy injection from ionizing radiation, leakage of UV, optical, or IR photons into the system, thermal heat transients, fluctuating externally-generated magnetic fields, and fluctuating externally-generated electric fields (e.g., RF). In these cases, the phenomena impinges on the qubit system \emph{and the immediate vicinity}, independent of the presence or absence of a qubit holding a quantum state. In these cases, we propose an appropriate sensor can potentially detect the error-inducing environmental disturbance without \emph{any} explicit or implicit influence on the state of a qubit in the vicinity of the disturbance. We henceforth refer to these error-inducing disturbances as ``environment''- or ``environmental''-inducing error sources. These errors are entirely incoherent errors within a computation.

A second class of error-inducing effects must also exist. This second class distinguishes itself through the quantum state entanglement produced as a result of the interaction between the error-inducing phenomenon and the presence of a qubit state. In a superconducting qubit device, we have in mind phenomena such as coupling to two-level state (TLS) systems and off-resonance coupling to other device elements. In these cases, a co-located measurement of the entangled error-inducing effect has the potential to produce back-action on the qubit's quantum state. Thus, we refer to these types of errors as ``entangling'' error sources. These ``entangling'' error-types can result in both incoherent and coherent error within computations.

We expect both types of errors described above are present in physical implementations of quantum computing systems. Throughout this study we have always assumed the entangling error is 60\% of the overall error probability.\footnote{Assuming 100\% of errors are of the entangling type is equivalent to the typical, pedagogical assumption in quantum error correction. Assuming 0\% of the errors are of the entangling type means \emph{all} errors are potentially identifiable by a co-located sensor, which we consider an unlikely and uninteresting, limiting case.}

\begin{figure*}[htb!]
\includegraphics[width=\textwidth]{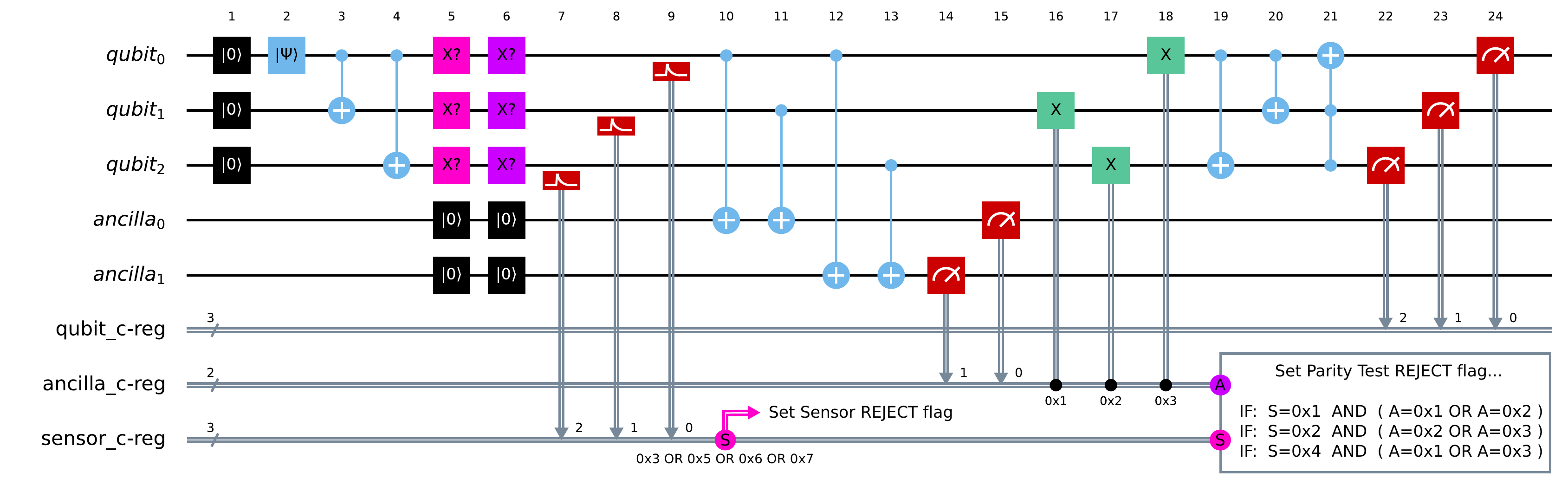}
\caption{\label{fig:QC-S111-E111-SingleCircuit}A quantum circuit for integrating sensor-assist into the standard 3-qubit, bit-flip quantum error correction code. The ``\textbf{\texttt{X}}?''-gates in columns 5~\&~6 represent possible  bit-flip (\textbf{\texttt{X}}-gate) errors on any of the computational qubits. See text for complete description.}
\end{figure*}

% Sensor assisted QEC
\subsection{\label{sec:middle_case}Sensor-assist in quantum error correction}

We now evaluate a more speculative scenario abstracted and generalized from the preceding sections. We assume all qubits experience \emph{entirely} independent errors and a co-located sensor is associated with each qubit. Furthermore, we assume a typical set of quantum computational gates is available and all errors in the error channel are bit-flip errors. Furthermore, we make the assumptions that circuit gates do not introduce errors outside of the error channel and ancilla qubits are reliable for their purpose of extracting a syndrome measurement. A number of such assumptions are made through-out the following development and these assumptions are explored in the Discussion (Sec.~\ref{sec:discussion}).

Figure~\ref{fig:QC-S111-E111-SingleCircuit} shows a quantum circuit for performing error correction when the error channel (columns 5~\&~6) is composed of independent environmental- and entangling-error types, as described above. In describing this quantum circuit, we focus on the key differences from a standard three-qubit, bit-flip error correction code. Columns~1-4 initialize three qubits, set a quantum state $|\Psi\rangle$ to preserve, and then encode the quantum state in the expanded three-qubit computational basis space. Column~5 includes a single bit-flip error (pink ``\textbf{\texttt{X}}?''-gates) on each of the three computational qubits, representing the potential environmental disturbance that can be detected by a co-located sensor. Column~6 represents the possibility to have entangling-type errors on any of the three qubits, shown as purple ``\textbf{\texttt{X}}?''-gates. Columns~7--9 represent the three co-located sensor readouts that are uniquely identified with each of the three physical qubits used for the state preservation.\footnote{Co-located sensors might also be associated with the ancilla qubits for further protection.} Note the diagram suggests the co-located sensors are near, but do not interact with the qubits. Pulses measured by the co-located sensors are recorded in the sensor's classical bit register, along the bottom of the diagram.

As the error correction portion of the circuit (columns 10--18) can only correct a single qubit error, at this point it is already possible to reject a single shot of the calculation if the co-located sensors measure two or more potential environmental disturbances to the qubits. When the sensor classical register reports 0x3, 0x5, 0x6, or 0x7, the Sensor REJECT flag is set for vetoing the calculation's output, as shown in the quantum circuit at column~10. Only in cases where a single (or no) co-located sensor has an event does the quantum computation fruitfully proceed to the error correction stage in columns~10--18. Assuming the calculation proceeds into the error correction stage in columns~10--18, the preserved quantum state is then decoded and measured in columns~19--24.

To understand the impact of the co-located sensor capacity to detect potential error-inducing environmental disturbances, we must evaluate the truth table of the circuit. There are eight possible combinations of errors for each of the environmental- and entangling-type errors (columns 5 and 6, respectively) on the three computational qubits, resulting in sixty-four possible error cases for the complete truth table (i.e., $2^3\times2^3=64$ error combinations). Note we are not yet invoking the assumption that the single error probability is ``small,'' though we will invariably evaluate specific cases under that assumption.

To compare the sensor-assisted circuit shown in Figure~\ref{fig:QC-S111-E111-SingleCircuit} to the standard, three-qubit, bit-flip error, quantum error correction code, recognize removal of columns 5, 7, 8, and 9 produces the standard three-qubit, bit-flip error correction circuit. Thus, we can tabulate the truth table for both circuits together for direct comparison. As stated above, there are 64 possible error combinations. The full 64 element truth table is provided in the Appendix.

\def\arraystretch{1.1}

\begin{table}[ht!]
    \small
    \centering
    \begin{tabular}{|cc|ccc|c|c|c|}
    \hline
\multicolumn{2}{|c|}{\textbf{Errors}} & \multicolumn{3}{c|}{\textbf{Gates}} & \textbf{Synd.} & \textbf{Prob.} & \textbf{Outcome} \\
\multicolumn{2}{|c|}{[Enviro.]} & \multicolumn{3}{c|}{Col. 5~\&~6} & Ancilla & Error $\times$ & Standard \\
\multicolumn{2}{|c|}{(Entangle)} & \multicolumn{3}{c|}{$\Rightarrow$ Result} & c-reg. & Non-error &  \textit{vs.} Assisted  \\
    \hline \hline
~[001]  & (000) & ~\textbf{\texttt{XI}}  &      & \textbf{\texttt{X}} &  &  $o^{1} \cdot p^{0}$ & \phantom{CC  \textit{vs.}  R$_{\mathrm{PT}}$}  \\
         &         & ~\textbf{\texttt{II}}  & ~$\Rightarrow$  & \textbf{\texttt{I}} & 0\texttt{x}3 & $\times$ &  C  \textit{vs.}  C \\
         &         & ~\textbf{\texttt{II}}  &      & \textbf{\texttt{I}} &  &  $\bar{o}^{2} \cdot \bar{p}^{3}$ &    \\ \hline
         & (001) & ~\textbf{\texttt{XX}}  &      & \textbf{\texttt{I}} &  &  $o^{1} \cdot p^{1}$ &   \\
         &         & ~\textbf{\texttt{II}}  & ~$\Rightarrow$  & \textbf{\texttt{I}} & 0\texttt{x}0 & $\times$ & CC  \textit{vs.}  CC  \\
         &         & ~\textbf{\texttt{II}}  &      & \textbf{\texttt{I}} &  &  $\bar{o}^{2} \cdot \bar{p}^{2}$ &    \\ \hline
         & (010) & ~\textbf{\texttt{XI}}  &      & \textbf{\texttt{X}} &  &  $o^{1} \cdot p^{1}$  &   \\
         &         & ~\textbf{\texttt{IX}}  & ~$\Rightarrow$  & \textbf{\texttt{X}} & 0\texttt{x}2 & $\times$ &  F  \textit{vs.}  R$_{\mathrm{PT}}$ \\
         &         & ~\textbf{\texttt{II}}  &      & \textbf{\texttt{I}} &  &  $\bar{o}^{2} \cdot \bar{p}^{2}$ &    \\ \hline
         & (100) & ~\textbf{\texttt{XI}}  &      & \textbf{\texttt{X}} &  &  $o^{1} \cdot p^{1}$  &   \\
         &         & ~\textbf{\texttt{II}}  & ~$\Rightarrow$  & \textbf{\texttt{I}} & 0\texttt{x}1 & $\times$ & F  \textit{vs.}  R$_{\mathrm{PT}}$  \\
         &         & ~\textbf{\texttt{IX}}  &      & \textbf{\texttt{X}} &  &  $\bar{o}^{2} \cdot \bar{p}^{2}$ &    \\ \hline
         & (011) & ~\textbf{\texttt{XX}}  &      & \textbf{\texttt{I}} &  &  $o^{1} \cdot p^{2}$ &   \\
         &         & ~\textbf{\texttt{IX}}  & ~$\Rightarrow$  & \textbf{\texttt{X}} & 0\texttt{x}1 & $\times$ & CC  \textit{vs.}  R$_{\mathrm{PT}}$  \\
         &         & ~\textbf{\texttt{II}}  &      & \textbf{\texttt{I}} &  &  $\bar{o}^{2} \cdot \bar{p}^{1}$ &    \\ \hline
         & (101) & ~\textbf{\texttt{XX}}  &      & \textbf{\texttt{I}} &  &  $o^{1} \cdot p^{2}$  &   \\
         &         & ~\textbf{\texttt{II}}  & ~$\Rightarrow$  & \textbf{\texttt{I}} & 0\texttt{x}2 & $\times$ &  CC  \textit{vs.}  R$_{\mathrm{PT}}$ \\
         &         & ~\textbf{\texttt{IX}}  &      & \textbf{\texttt{X}} &  &  $\bar{o}^{2} \cdot \bar{p}^{1}$ &    \\ \hline
         & (110) & ~\textbf{\texttt{XI}}  &      & \textbf{\texttt{X}} &  &  $o^{1} \cdot p^{2}$  &   \\
         &         & ~\textbf{\texttt{IX}}  & ~$\Rightarrow$  & \textbf{\texttt{X}} & 0\texttt{x}0 & $\times$ & F  \textit{vs.}  F  \\
         &         & ~\textbf{\texttt{IX}}  &      & \textbf{\texttt{X}} &  &  $\bar{o}^{2} \cdot \bar{p}^{1}$ &    \\ \hline
         & (111) & ~\textbf{\texttt{XX}}  &      & \textbf{\texttt{I}} &  &  $o^{1} \cdot p^{3}$  &   \\
         &         & ~\textbf{\texttt{IX}}  & ~$\Rightarrow$  & \textbf{\texttt{X}} & 0\texttt{x}3 & $\times$ &  F  \textit{vs.}  F \\
         &         & ~\textbf{\texttt{IX}}  &      & \textbf{\texttt{X}} &  &  $\bar{o}^{2} \cdot \bar{p}^{0}$ &    \\ \hline
    \end{tabular}
    \caption{Truth table for the case when a single environmental-type error occurs on qubit~0 (i.e., error mask: $[001]$), with any combination of entangling-type errors (i.e., error masks: $(000)$-$(111)$~). Outcome notation: C = Correct, CC = Correct via cancellation, F = Faulty, and R$_{\mathrm{PT}}$ = REJECT based on syndrome parity test. See text for complete table description.}
    \label{tab:001_cases}
\end{table}

\def\arraystretch{1.25}

We focus on the interesting case when there is a \emph{single} qubit affected by an environmental-type disturbance phenomenon (in column 5), detectable by a co-located sensor. We assume 100\% of environmental-type phenomena are detected by the co-located sensors, though this is not required for gaining utility from a sensor-assist method. The \textbf{Outcome} column of Table~\ref{tab:001_cases} presents the eight outcome cases when a single environmental disturbance occurs on \textit{qubit}$_0$, with any possible combination of entangling errors on the three qubits. An error bit-mask notation is used to uniquely identify each possible error case. For example, in our bit-mask notation [001]~(011) means an environmental disturbance has caused a bit-flip error on \textit{qubit}$_0$ and two entangling-type bit-flip errors have occurred on \textit{qubit}$_0$ and \textit{qubit}$_1$, the \textit{qubit} designations referring again to the quantum circuit in Figure~\ref{fig:QC-S111-E111-SingleCircuit}. This bit-mask notation is given in the \textbf{Errors} column of Table~\ref{tab:001_cases}.

Note we are \emph{not} assuming that only a single error occurs in the error channel. We take for granted that if the error probabilities are ``small,'' then the probability of multiple errors occurring will diminish greatly. For additional clarity, in addition to the error bit-mask identifiers, the \textbf{Gates} column in Table~\ref{tab:001_cases} presents the quantum gates for both types of errors, represented in columns 5~\&~6 in the quantum circuit (Fig.~\ref{fig:QC-S111-E111-SingleCircuit}). The assumption used in this report that all errors are bit-flip errors has an unintended consequence that two bit-flip errors can cancel if they both appear on the same qubit. Thus, the resultant gate for each of the three qubits is presented in Table~\ref{tab:001_cases}, where \textbf{\texttt{X}} is a bit-flip error gate and \textbf{\texttt{I}} is the identity gate (i.e., no error). This cancellation effect is an artifact of the unrealistic model of pure bit-flip errors.

For each error combination, the \textbf{Synd.} column in Table~\ref{tab:001_cases} provides the syndrome measurement (columns 10--15 in Fig.~\ref{fig:QC-S111-E111-SingleCircuit}) recorded in the ancilla classical register. In the lower right of Figure~\ref{fig:QC-S111-E111-SingleCircuit}, classical logic is used to assess if the combination of the co-located sensor and the parity tests performed in the syndrome measurement are consistent with a single error on the qubit associated with the co-located sensor reporting an environmental disturbance. Each unique error combination results in a specific outcome from the quantum circuit. If no errors of any kind occur, then the circuit returns the correct (C) quantum state. Likewise, if only a single error occurs of either type (environmental or entangling), again the circuit returns the correct outcome (C). In some cases, as we have mentioned, the bit-flip error induced by the environmental disturbance is canceled by an entangling error on the same qubit. In these cases, such as [001]~(001), the quantum circuit returns the correct outcome quantum state, but via a fortuitous cancellation, a ``correct via cancellation'' (CC) outcome state.

As the number of error occurrences in the error channel increase, the standard error correction code and the sensor-assisted code return different outcomes. This is the first notable conclusion: The sensor-assisted code only has an impact for cases when the quantum state has an uncorrectable error. In this way one intuits correctly that the classical information provided by a co-located sensor can't increase the number of correctly returned quantum states. However, the sensor-assist method can identify when an uncorrectable error has likely occurred, giving the user the opportunity to remove the calculation from further consideration in a computational effort.

To quantify these statements, we define several error probability notation terms, used in part in the probability (\textbf{Prob.}) column in Table~\ref{tab:001_cases}. In this column, $o$ is the probability of an environmentally-induced error and $p$ is the probability of an entangling-type error. The non-error complements are $\bar{o}=1-o$ and $\bar{p}=1-p$. We also define $\hat{P}=o+p-op$, the probability that at least one error occurred in the error channel (i.e., the combination of columns~5~\&~6 in Fig.~\ref{fig:QC-S111-E111-SingleCircuit}). Note $\hat{P}$ is \emph{not} the probability that a qubit is in an error state after the error channel gates have been applied (i.e., the combined action of columns~5~\&~6 in Fig.~\ref{fig:QC-S111-E111-SingleCircuit}). That is, $\hat{P}$ does not correspond to what one would measure as a qubit error rate except when the error is only either 100\% entangling-type or 100\% environmental-type. See the Appendix for a full derivation and definition of the terms $\hat{P}$, $o$, $p$, $\bar{o}$, and $\bar{p}$.

Looking again at Table~\ref{tab:001_cases}, if two or more \textbf{\texttt{X}}-gates appear in the resultant gate column, the standard bit-flip quantum error correction code will run to completion, but the returned quantum state will be faulty (F). The sensor-assisted method, however, is able to set a Parity Test REJECT flag (R$_{\mathrm{PT}}$) in half of the faulty cases.\footnote{Note the probability of these multiple error cases occurring in a real set of calculations is not half. That is, there are 64 possible error cases, but there is not equal probability weight of arriving at each of the 64 error cases for a set of calculations.} The computational advantage of the sensor-assist method comes from the fact that the Parity Test REJECT flag is set for cases when the number of ``small'' probability errors is low. To see this, consider the probability (\textbf{Prob.}) column in Table~\ref{tab:001_cases}, which shows as an exponent the number of errors occurring. The sum of the exponents of the $o$ and $p$ terms reveals the \emph{order} of ``small'' probability errors. By examining Table~\ref{tab:001_cases}, it is possible to see that whereas the standard error correction code permits faulty computations to pass through at an error-order of $2$ and higher, the sensor-assist code will only allow faulty computations to pass through at an error-order of $3$ or higher. This computational benefit does, however, come at the expense of also rejecting correct computations at an error-order of $3$ that are arrived at through fortuitous cancellations (CC). As a reminder, the fortuitous cancellation (CC) cases, are artifacts of the simplistic model of treating all errors as single bit-flips.

\begin{table*}[ht!]
    %\small
    \centering
    \begin{tabular}{l|r|r|r|r|r|r|r|r|}
     \multicolumn{1}{l}{} & \multicolumn{8}{c}{\textbf{Outcome fractions, $\mathcal{F}$, for various error probabilities, $\hat{P}$ and $p$}} \\
    \hline
     & \multicolumn{2}{c|}{$\hat{P}=0.20$, $p=0.20$} & \multicolumn{2}{c|}{$\hat{P}=0.20$, $p=0.12$}  & \multicolumn{2}{c|}{$\hat{P}=0.05$, $p=0.03$} & \multicolumn{2}{c|}{$\hat{P}=0.05$, $p=0.01$} \\
    Outcome case & Standard & Assisted & Standard & Assisted & Standard & Assisted & Standard & Assisted \\
    \hline
    Correct (C)                             &\ 0.8960\ &\ 0.8960\ &\ 0.8751\ &\ 0.8751\ &\ 0.9911\ &\ 0.9911\ &\ 0.9917\ &\ 0.9917\ \\
    Correct via cancellation (CC)           &\ 0.0000\ &\ 0.0000\ &\ 0.0312\ &\ 0.0209\ &\ 0.0018\ &\ 0.0017\ &\ 0.0012\ &\ 0.0011\ \\
    Faulty (F)                              &\ 0.1040\ &\ 0.1040\ &\ 0.0937\ &\ 0.0331\ &\ 0.0071\ &\ 0.0025\ &\ 0.0071\ &\ 0.0003\ \\ 
    Parity Test REJECT (R$_{\mathrm{PT}}$)  &\ -\      &\ 0.0000\ &\ -\      &\ 0.0476\ &\ -\      &\ 0.0034\ &\ -\      &\ 0.0022\ \\
    Sensor REJECT (R$_{\mathrm{S}}$)        &\ -\      &\ 0.0000\ &\ -\      &\ 0.0233\ &\ -\      &\ 0.0013\ &\ -\      &\ 0.0047\ \\
    \hline
    Effective correct outcome $\rightarrow$ &\ 0.8960\ &\ 0.8960\ &\ 0.9063\ &\ 0.9644\ &\ 0.9929\ &\ 0.9974\   &\ 0.9929\   &\ 0.9997\ \\
    \hline
    \end{tabular}
    \caption{Standard bit-flip quantum error correction outcome fractions compared to those from the sensor-assisted quantum circuit. Here $\hat{P}=o+p-op$ is the probability for any error to occur in the error channel (i.e., the combined columns~5~\&~6 in Fig.~\ref{fig:QC-S111-E111-SingleCircuit}). The entangling-type error (i.e., column~6 in Fig.~\ref{fig:QC-S111-E111-SingleCircuit}) has probability $p$ and the sensor detectable environmental disturbance-induced error (i.e., column~5 in Fig.~\ref{fig:QC-S111-E111-SingleCircuit}) has probability $o$. See the text body for further details and the Appendix for the derivation of the relationship between $\hat{P}$, $o$, and $p$.}
    \label{tab:probabilities}
\end{table*}

Finally, Table~\ref{tab:probabilities} presents numerical values for several specific choices of error probability, parameterized by $\hat{P}$ and the entanglement error $p$. The values of the error probabilities are merely illustrative. There are four examples, and for each example, the standard bit-flip quantum error correction code is compared to the sensor-assisted code. We present the fractional weights of specific outcomes from the quantum circuit in Figure~\ref{fig:QC-S111-E111-SingleCircuit}, as described above in the explanation of Table~\ref{tab:001_cases}, with the addition of the Sensor REJECT (R$_{\mathrm{S}}$) cases (which appear in the full 64-combination tables in the Appendix).

From Table~\ref{tab:probabilities} we see several features. First, the fractional weight for the correct outcome (C) is always the same for the standard code and the sensor-assisted code, the ``intuition'' mentioned above. Second, when $\hat{P}=0.20$ and $p=0.20$, the environmental disturbance error probability is zero, so the two codes perform the same. Third, the key metric for determining the computational advantage is effective correct outcome fractional weight calculated as $(C+CC)/(C+CC+F)$. As a portion of the cases are removed from consideration by the logic of the sensor-assisted method, the denominator is lower than for the standard quantum correction code. The case $\hat{P}=0.20$ and $p=0.12$ is quoted in the abstract of this report. Fourth, as the overall scale of the error's fractional weighting decreases, the utility of the sensor-assist method decreases, as one would also intuitively expect.

%%
%% DISCUSSION
%%

\section{\label{sec:discussion}Discussion}

A number of assumptions were made in the foregoing analysis. It is valuable to explore the limitations these assumptions may impose on the results of this work. First and foremost, we assumed all quantum computation error types are of the bit-flip variety. In the case of using a co-located sensor for simply ``vetoing'' selected calculations in response to the detection of an environmental disturbance (Sec.~\ref{sec:example_error}), this choice of error-type is of no substantive consequence since any error type is still subject to the same ``veto'' of the entire computation. However, one might argue the errors present in the actual Yorktown backend calculation are not even discrete in nature. That is, our assumption of a bit-flip type error is effectively assuming the co-located sensor is responding to discrete events, like the interaction of an ionizing $\gamma$-ray in the chip substrate. If the environmental disturbances are of a continuous nature, it may be difficult to know when the co-located sensor is reporting a disturbance warranting rejection of the calculation instance. This could be assessed through empirical correlation studies to determine at what level of co-located sensor response it becomes beneficial to reject a specific calculation.

Perhaps more pointedly, even the standard textbook example three-qubit, bit-flip quantum error correction scheme \emph{presumes knowledge} of the error type. In other words our assumption of a bit-flip error type is entirely analogous to pedagogical presentations~\cite{10.5555/1972505} of a purely quantum method of error correction. We believe a key point is that if a co-located sensor's response is \emph{preferentially correlated} with a specific type of correctable error in the quantum calculation, then a sensor-assisted mitigation code implementation is likely fruitful. Furthermore, while not shown in this report, the quantum circuit developed in this report for use with co-located sensors also works for phase-flip errors when Hadamard gates are inserted on each computational qubit at what would be columns 4.5 and 9.5 in Figure~\ref{fig:QC-S111-E111-SingleCircuit}, as well as changing the error types in columns 5 and 6 to \textbf{\texttt{Z}}-gates.

Related to the exclusive use of bit-flip type errors in this report's analysis is the, as we have called them, ``fortuitous cancellations'' that arise as a natural (logical) consequence to the introduction of two independent error types within the error channel. We would readily agree with the reader that it seems highly unlikely that two such errors, of presumably very different phenomenological cause, would perfectly cancel each other on a single qubit. The specific case of concern would need evaluation in the framework of Table~\ref{tab:001_cases}.

In reality, the type of ``errors'' induced by ionizing radiation interactions in superconducting qubit devices is not entirely unknown. Our prior work~\cite{Oliver2020} has shown elevated levels of ionizing radiation results in increased quasiparticle density in the qubits' superconducting circuits. As quasiparticles tunnel through the Josephson junctions of, for example, transmon qubits, the parity of the quantum state flips. Thus, the appearance of parity transitions in transmon qubits due to tunneling of quasiparticles~\cite{ISI:000320589900109,PhysRevApplied.12.014052,PhysRevB.84.064517,PhysRevLett.121.157701} is a signature of energy injections due to ionizing radiation. The transitions rates of qubit relaxation and dephasing due to quasiparticle tunneling through Josephson junctions was previously investigated~\cite{ISI:000320589900109}.

In this report, we have presented simple methods of sensor-assisted fault mitigation in quantum computation. We anticipate sensor-assisted fault mitigation is possible within the frameworks of surface and stabilizer codes, though we have not explored those possibilities in any detail. Surface codes are potentially particularly interesting as it is easy to envision a physical surface array of single-qubit transmon chips, each containing a QET sensor. A high-quality chip-to-chip communication method would need development, but it is perhaps achievable through air-bridges or capacitive coupling elements in the circuits.

In this report, we have consistently had in mind ionizing radiation as representative of a class of environmental disturbing effects to superconducting transmon qubit systems. We proposed a specific sensor type---the QET---as a means for detecting these ionizing radiation specific environmental disturbances. At the present time, ionizing radiation is a minor contributor to quantum computational error. However, we note plans for future quantum computing systems, such as the ``Goldeneye'' million-qubit capable cryostat IBM is building~\cite{ScienceNews-Goldeneye}, are reaching the same physical scale as deep underground cryogenic research instruments~\cite{ALDUINO20199} that actively, passively, and in analysis work against ionizing radiation as a background to their experimental detection goals. The likelihood of ionizing radiation interactions occurring increases roughly linearly with the mass of the instrument, the total silicon chip substrate mass in the case of transmon qubit. Once the extraneous silicon chip substrate mass is minimized, the interaction likelihood of ionizing radiation within a single computational cycle will scale directly with the number of qubits (and duration of the computation). In this regime of large-scale qubit systems (and long duration computations) we believe the utility of sensor-assisted fault mitigation is likely to grow.

A key question is whether these considerations extend beyond ionizing radiation to other, more general, environmental disturbances. We believe the QET co-located sensor approach described in this report is applicable to most silicon chip-based superconducting Josephson junction qubit devices (e.g., flux, charge, and phase qubit varieties). However, the broader objective of the analysis presented in this report was to show the potential computational value achievable \emph{if} quantum computational error types are preferentially correlated with sensor-detectable environmental disturbances. For superconducting transmon qubit systems, other case types may include IR photon leakage sensing, vibration-induced energy coupling, and stray electric- or magnetic-field fluctuations. We are not in a position to speculate on analogous environmental disturbance error types and sensor combinations in other qubit modalities. We look to experts in the relevant disciplines to consider if the ideas presented in this report are transferable to other quantum computing systems.

During the final preparation of this report for submission, we were made aware of an article by J.M.~Martinis~\cite{martinis2020saving} which presented a model of ionizing radiation induced errors in superconducting qubits. Of interest to our own report, the Martinis article touches on error correction in the face of disturbances from ionizing radiation. In particular, Martinis states, ``if errors are large or correlated, bunching together either in time or across the chip in space, then error decoding fails.'' We concur with this assessment as it relates to disturbances from ionizing radiation and find the design solutions suggested by Martinis to be compelling. Our own suggested design solution, described in this report, is to place uniquely paired sets of a qubit and a sensor together on shared chip substrate. Communication via air-bridges, capacitive coupling, or other novel means for qubit-to-qubit interconnection is required to create a network of qubits for computation. In this way, we see no inconsistencies between the concepts presented in this report and the concepts presented by Martinis.

%%
%% SUMMARY
%%

\section{\label{sec:summary}Summary}

In this report, we proposed hybrid superconducting device concepts for quantum computation. The inclusion of a co-located sensor on qubit substrates provides the potential to detect environmental disturbances causing errors in a quantum computation. In the simplest form, such co-located sensors provide a means to selectively ``veto'' and reject just those calculations where an environmental disturbance is likely to result in an incorrect calculation result. We showed the computational advantage of such a scheme and proposed device concepts that could implement such error mitigating techniques using proven device fabrication designs and methods.

We abstracted the co-located sensor concept to a scenario where every qubit has a uniquely assigned co-located sensor. We developed a formulation of the three-qubit, bit-flip quantum error correction code to take advantage of the co-located sensor's ability to detect environmental disturbances. The results demonstrated an enhanced effective quantum computational performance at the cost of the rejection of some calculation repetitions.

In both fault mitigation concepts considered in this report, the computational enhancements are numerically modest. Nevertheless, we believe these results recommend the development and investigation of a new class of superconducting quantum computation devices that include co-located sensors for the detection of environmental disturbances. We believe such devices are a potential new tool in the broad category of hybrid quantum-classical algorithm development and approaches to quantum error mitigation~\cite{endo2020hybrid}.

\begin{acknowledgments}
The concepts presented in this work stem from efforts underway at Pacific Northwest National Laboratory (PNNL). PNNL is a multiprogram national laboratory operated by Battelle for the U.S. Department of Energy under Contract No. DE-AC05-76RL01830. The research results presented in this work were developed outside of any supporting grant or internal R\&D support mechanism and are associated with U.S. provisional patent application number~63/123,050 (9 December 2020). The authors thank Alexander Melville and Kyle Serniak (both at MIT Lincoln Laboratory) for answering questions regarding how superconducting qubits located on separate chip substrates might inter-communicate through superconducting air-bridges or capacitive coupling across gaps between chips, making some of the speculative device concepts we propose seem more plausible to the authors. We acknowledge the use of IBM Quantum services for this work. The views expressed are those of the authors, and do not reflect the official policy or position of IBM or the IBM Quantum team. The authors thank Mark V. Raugas and Tobias J. Hagge (both at PNNL) for constructive comments on an early draft of this report. PNNL Information Release PNNL-SA-158581.

%Example backend citation: ibmq\_vigo v1.0.2, IBM Quantum team. Retrieved from \url{https://quantum-computing.ibm.com} (2020)

\end{acknowledgments}

%\cite{*[{prepended text}][{appended text}]key}

%\bibliography{references}
\input{main.bbl}

%%
%% Appendix
%%

\appendix

\begin{figure}[b!]
    \centering
    \includegraphics[width=\columnwidth]{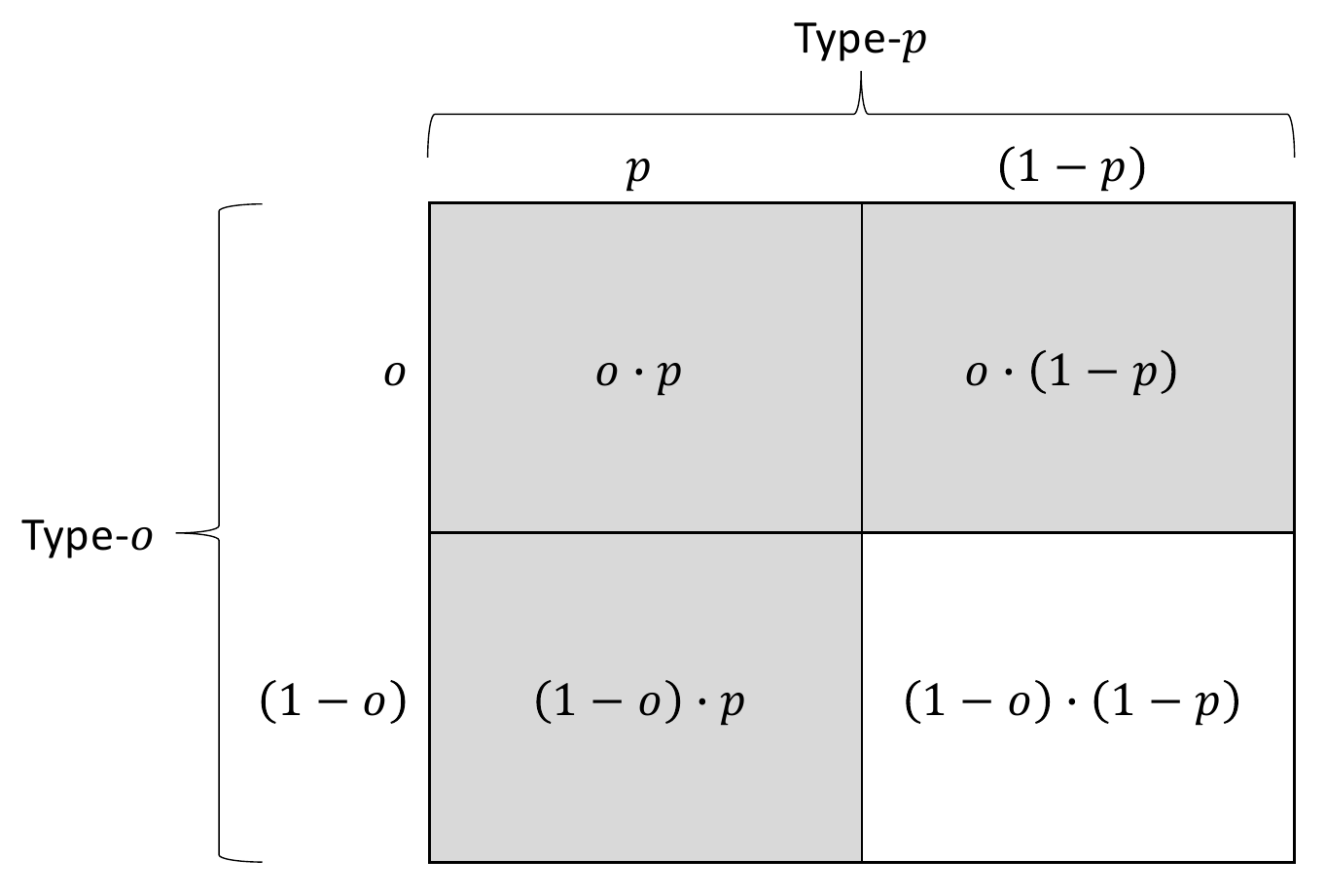}
    \caption{The possibilities for combining two independent random event types with individual probabilities of occurrence, $o$ and $p$. The values $(1-o)$ and $(1-p)$ are the individual probabilities for each random event type to \emph{not} occur. The shaded boxes represent when at least one event occurs.}
    \label{fig:outcome_table}
\end{figure}

% Other concepts
\section{Other sensor-assisted qubit concepts}

\begin{figure*}[htb!]
\begin{subfigure}{.32\textwidth}
  \centering
  \includegraphics[width=\textwidth]{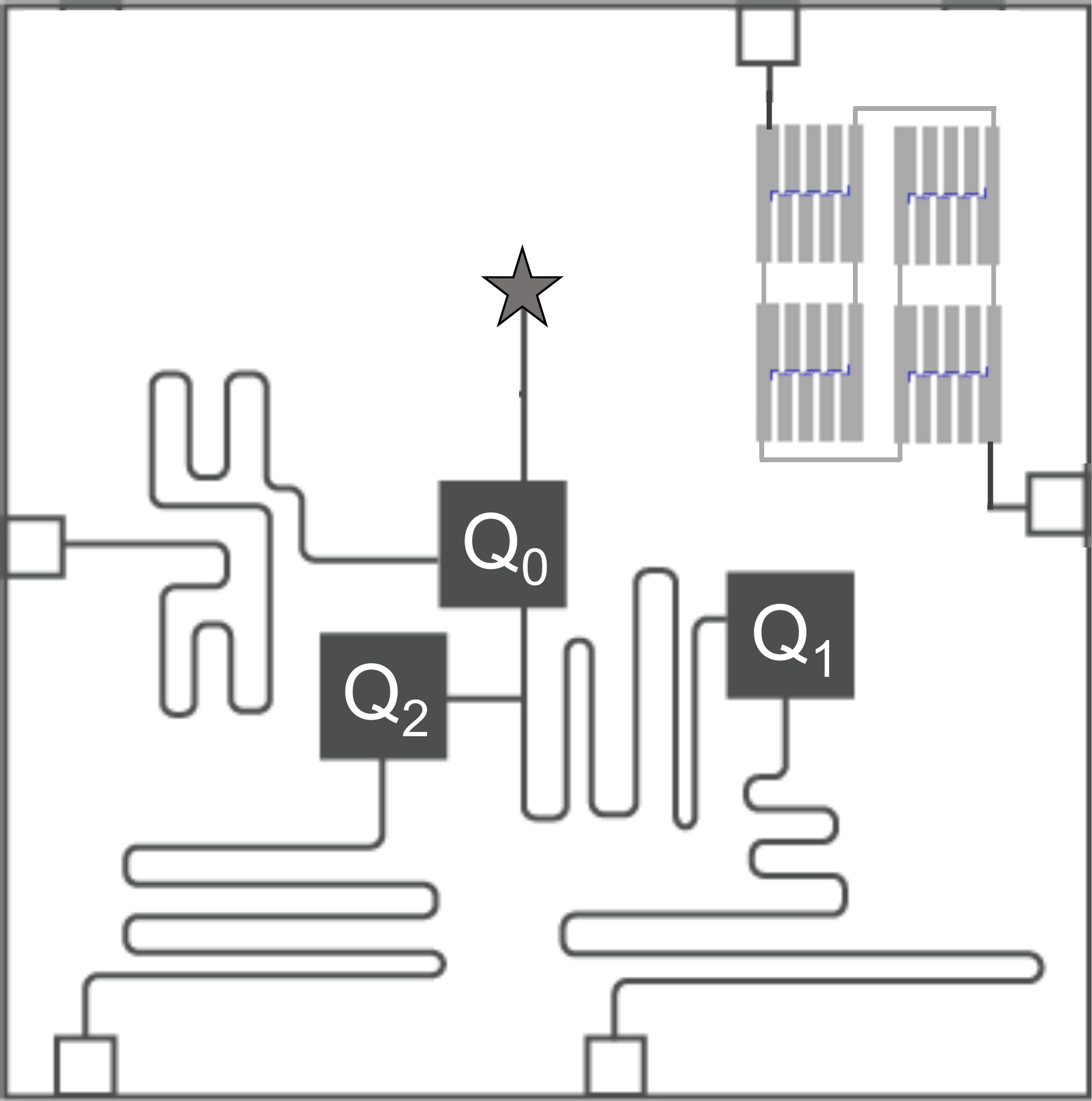}
  \caption{Qubit and QET sensor concept.}
  \label{fig:3-qubit-QET-scheme}
\end{subfigure}
\begin{subfigure}{.32\textwidth}
  \centering
  \includegraphics[width=\textwidth]{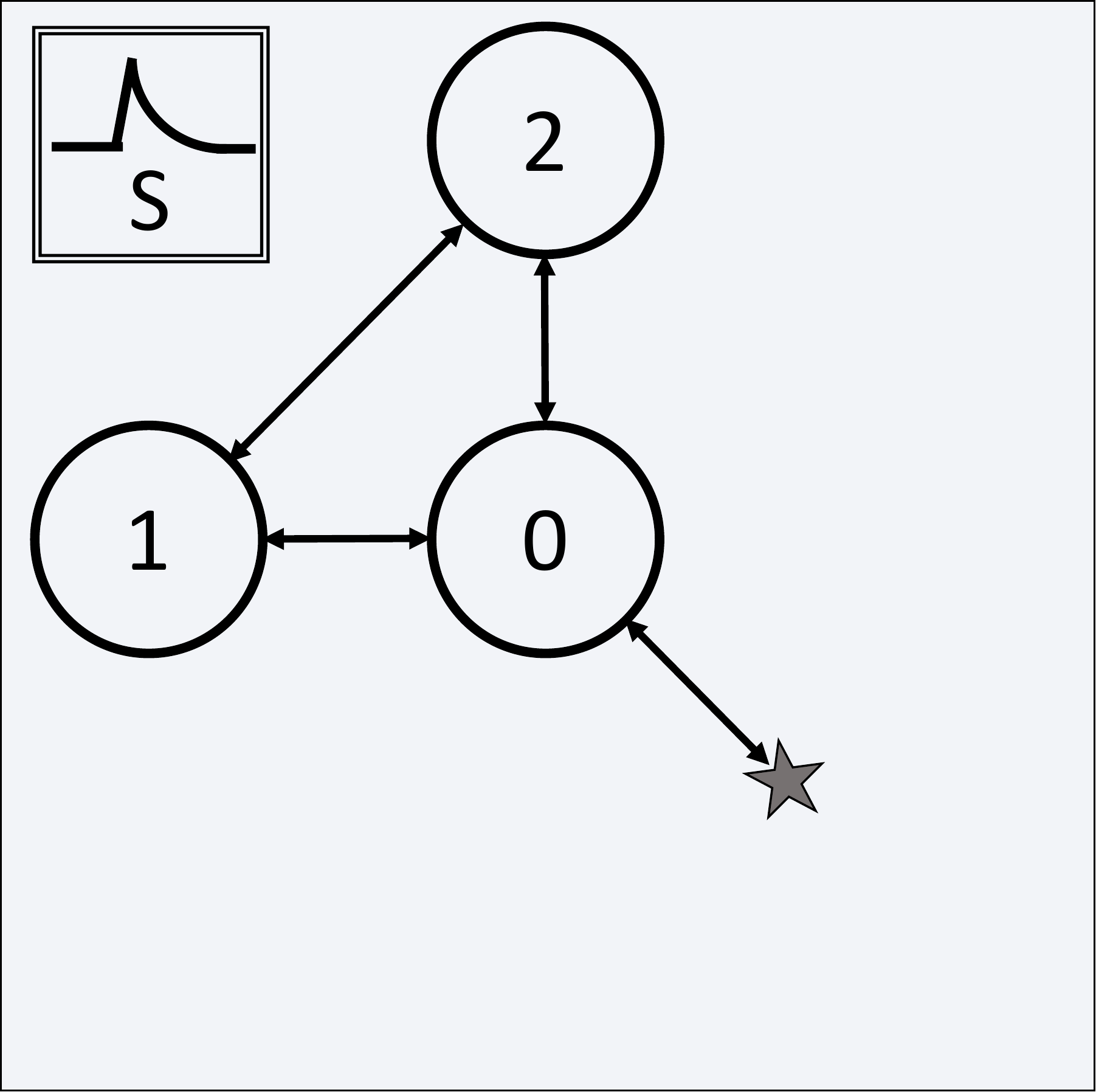}
  \caption{Scheme for 3 qubit and QET.}
  \label{fig:3-qubit-QET-connections}
\end{subfigure}
\begin{subfigure}{.29\textwidth}
  \centering
  \includegraphics[width=\textwidth]{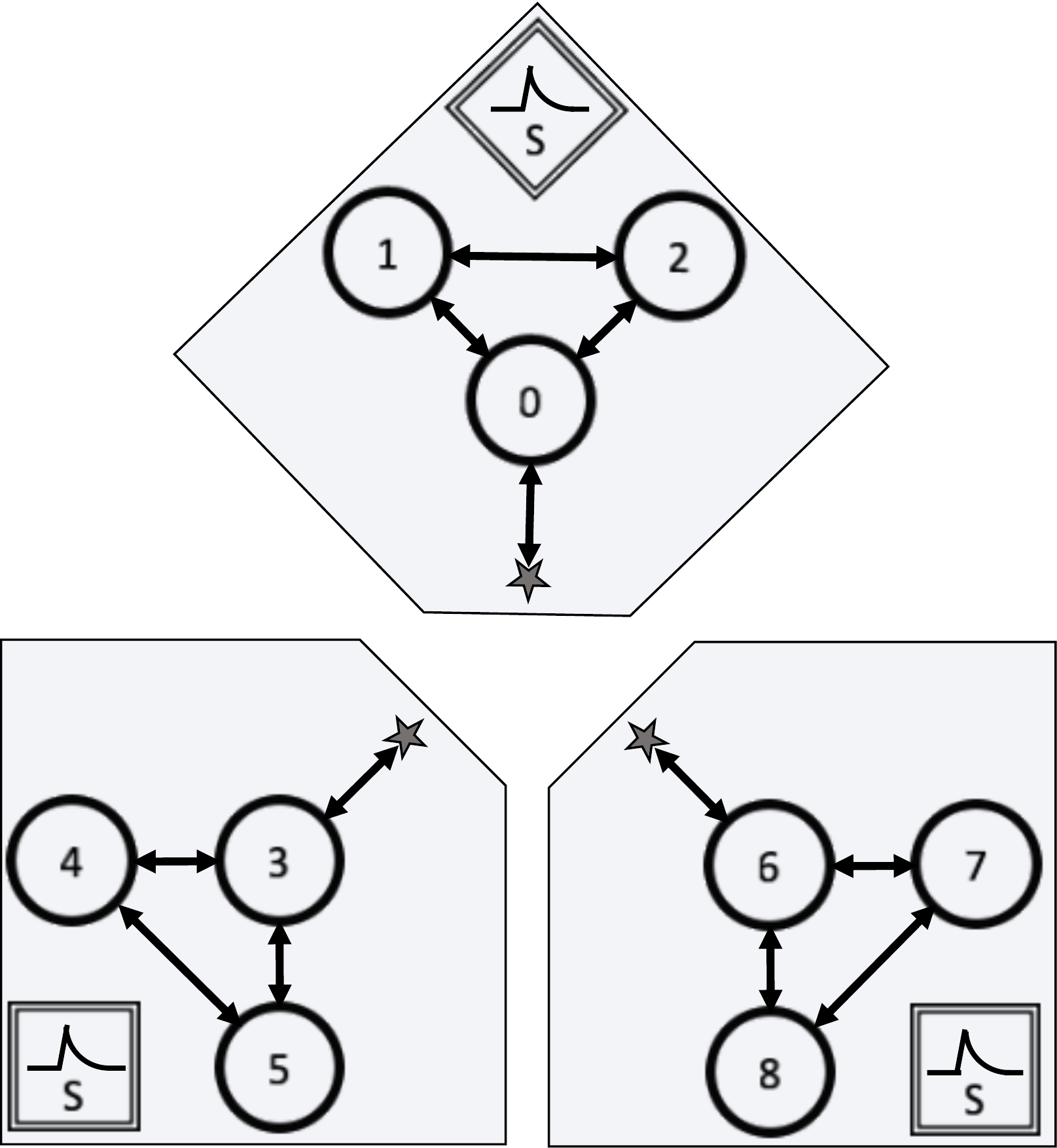}
  \caption{Three chips of 3 qubits each.}
  \label{fig:3-chip-3-qubit}
\end{subfigure}
\caption{Concepts for 3-qubit devices utilizing co-located QET sensors. The star symbol, $\star$, represents a chip-to-chip inter-communication point.}
\label{fig:multi-qubit-TES-scheme}
\end{figure*}

\begin{figure*}[htb!]
  \centering
  \includegraphics[width=\textwidth]{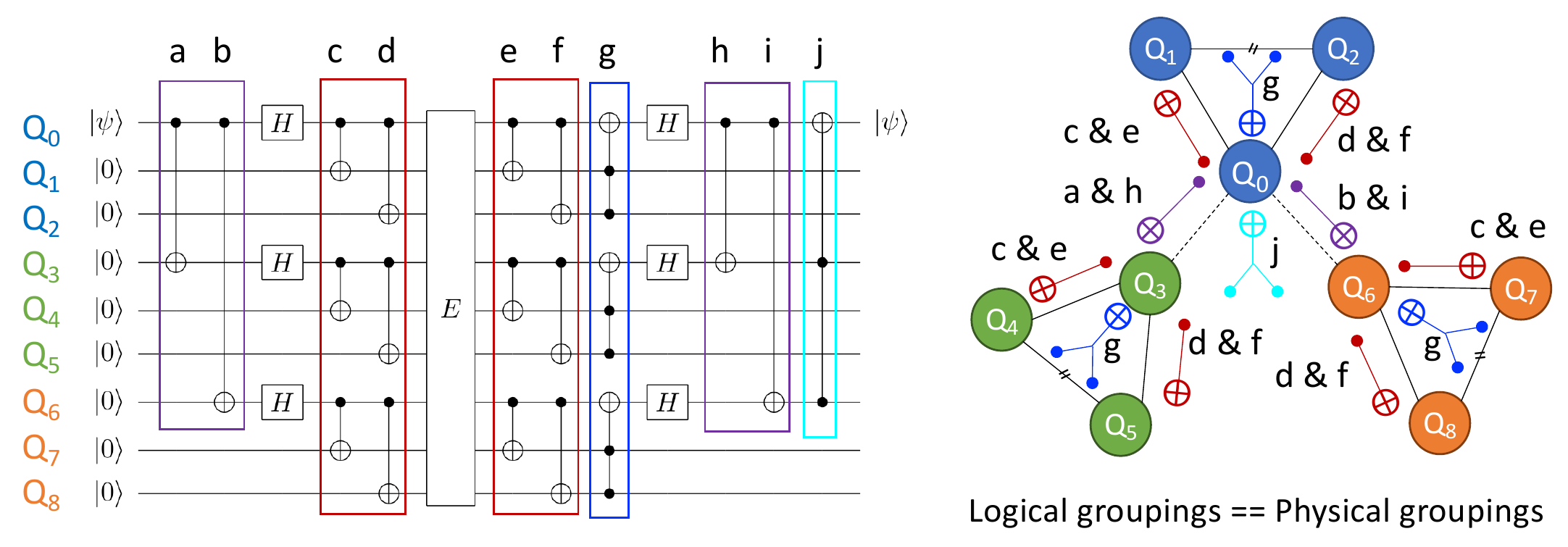}
\caption{The 9-qubit Shor code, left, with qubit groupings and alpha labels on the computational steps. To the right, qubit groupings with a physical arrangement similar to that shown in Figure~\ref{fig:3-chip-3-qubit}, with the qubit computation gates displayed. The thin black lines connecting the qubit triplets represent the on-chip connections between the qubit triplet, each triplet color coded blue, green, and orange.}
\label{fig:3x3-physical}
\end{figure*}

\begin{figure*}[htb!]
  \centering
  \includegraphics[width=\textwidth]{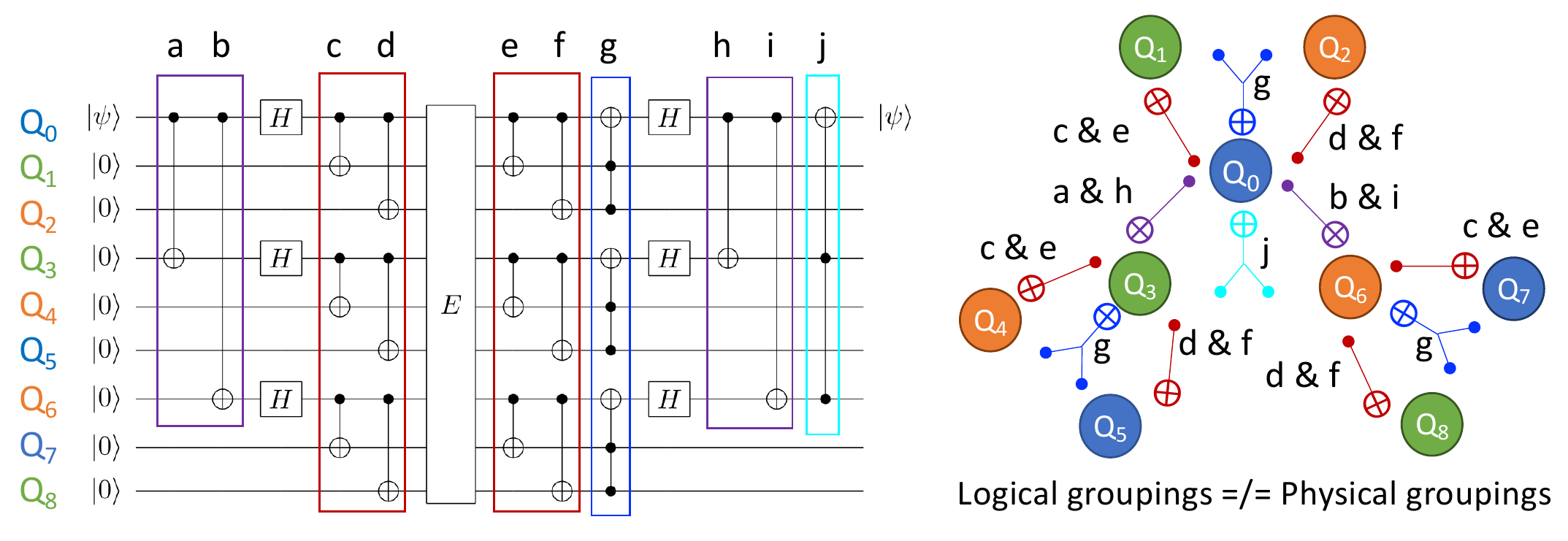}
\caption{Similar to Fig~\ref{fig:3x3-physical}; however, now the physical qubits are distributed throughout the Shor code. In this case, shown on the right, the matching color qubit triplets are each on a single chip substrate.}
\label{fig:3x3-logical}
\end{figure*}

Early in the development of the sensor-assisted quantum error correction concepts presented in this report, we explored integrating co-located sensors into the 9-qubit Shor code. We envisioned groups of qubits on shared substrates, monitored by co-located QET sensors; see Figure~\ref{fig:multi-qubit-TES-scheme}. We envisioned 3-qubit chips, in a grouping of three chips, to provide the 9~physical qubits needed for the Shor code. The concept would assume a standard Toffoli gate (CCNOT gate) implementation is available and that there is a means for interconnecting the three chips. Concepts for physical implementation in superconducting Josephson multi-qubit devices were recently explored in the literature~\cite{PhysRevA.101.022308}, and we believe chip-to-chip air bridges or capacitive coupling are future possibilities. Use of ancilla qubits is also likely in a practical implementation, though that was not considered in these initial concepts.

The 9-qubit Shor code contains several 3-fold symmetries we believed would prove advantageous for using co-located sensors to provide informed error correction coding. Figures~\ref{fig:3x3-physical} and~\ref{fig:3x3-logical} present these ideas. In each of the figures, the computational gates are assigned a designating letter (a--j) and are grouped within colored boxes. The qubits residing on the same chip share the same color (blue, green, or orange).

The qubits can be grouped in a set of three so that the chip-to-chip communication is minimized (Fig.~\ref{fig:3x3-physical}). However, one sees the signals from a co-located sensor will flag an entire sub-group of the qubits as potentially error prone, making the Shor code fail, in general. An alternative is to distribute the physical qubits across the Shor code (see Fig.~\ref{fig:3x3-logical}) but at the expense of having the majority of the multiple qubit gates require chip-to-chip communication. Worse, one sees once again a single co-located sensor event flags three qubits across the Shor code as potentially being in error. As the Shor code, in general, can only protect against two qubit errors, it was realized this approach was likely not fruitful.

From this analysis, we abandoned further development of error correction where a co-located sensor is assigned to more than one single (independent) qubit. However, we expect for specific computation implementations, there may yet be utility in considering symmetries within the computation to determine how to efficiently arrange co-located sensors whilst minimizing error prone qubit-to-qubit inter-communications.

% Repetition calculation statistics
\section{Statistics of a repeated calculation}

\begin{figure*}[t]
%%%\centering
%%%\raggedright
%%%\begin{center}
%%%    \underline{\textbf{Balanced Deutsch-Jozsa calculation results - I}} \\
%%%    \vspace{-0.6em}
%%%\end{center}
(a)%%%~\vspace{-1.2em}
%%%\begin{center}
\begin{subfigure}{0.95\columnwidth}
  %\centering
  \includegraphics[width=\columnwidth,center]{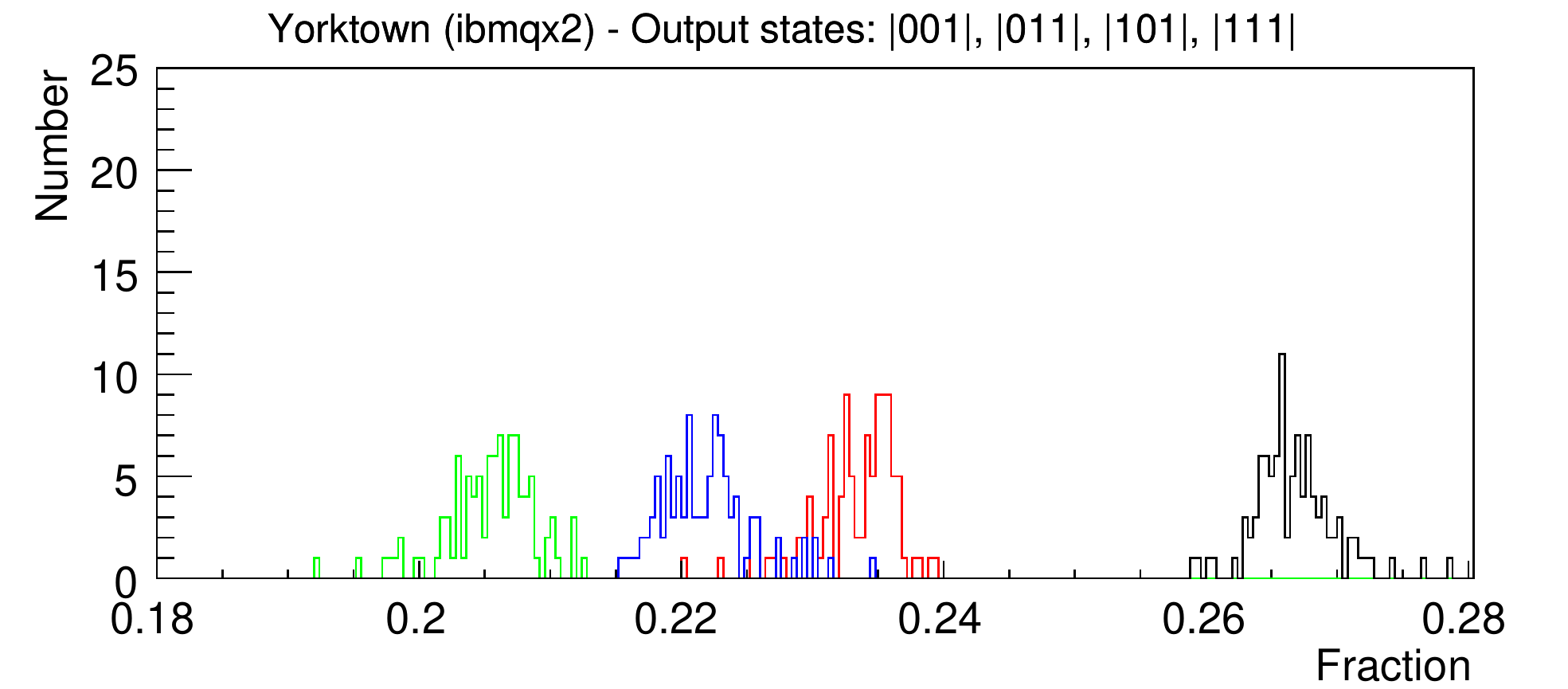}
\end{subfigure}
%%%\vspace{-1.2em}
%%%\end{center}
%\phantom{---} \\
\quad (e)%%%~\vspace{-1.2em}
%%%\begin{center}
\begin{subfigure}{0.95\columnwidth}
  %\centering
  \includegraphics[width=\columnwidth,center]{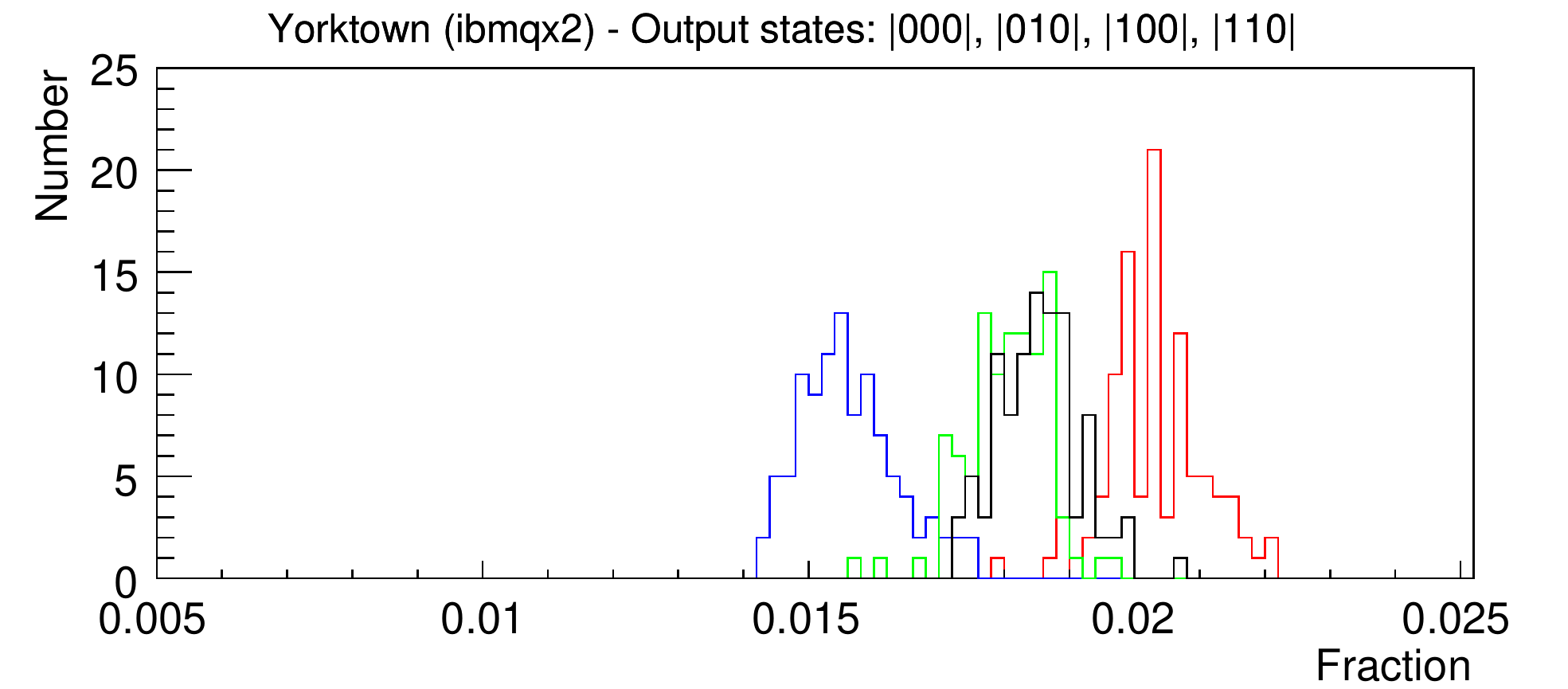}
\end{subfigure}
%%%\vspace{-1.2em}
%%%\end{center}
%\phantom{---} \\
(b)%%%~\vspace{-1.2em}
%%%\begin{center}
\begin{subfigure}{0.95\columnwidth}
  %\centering
  \includegraphics[width=\columnwidth,center]{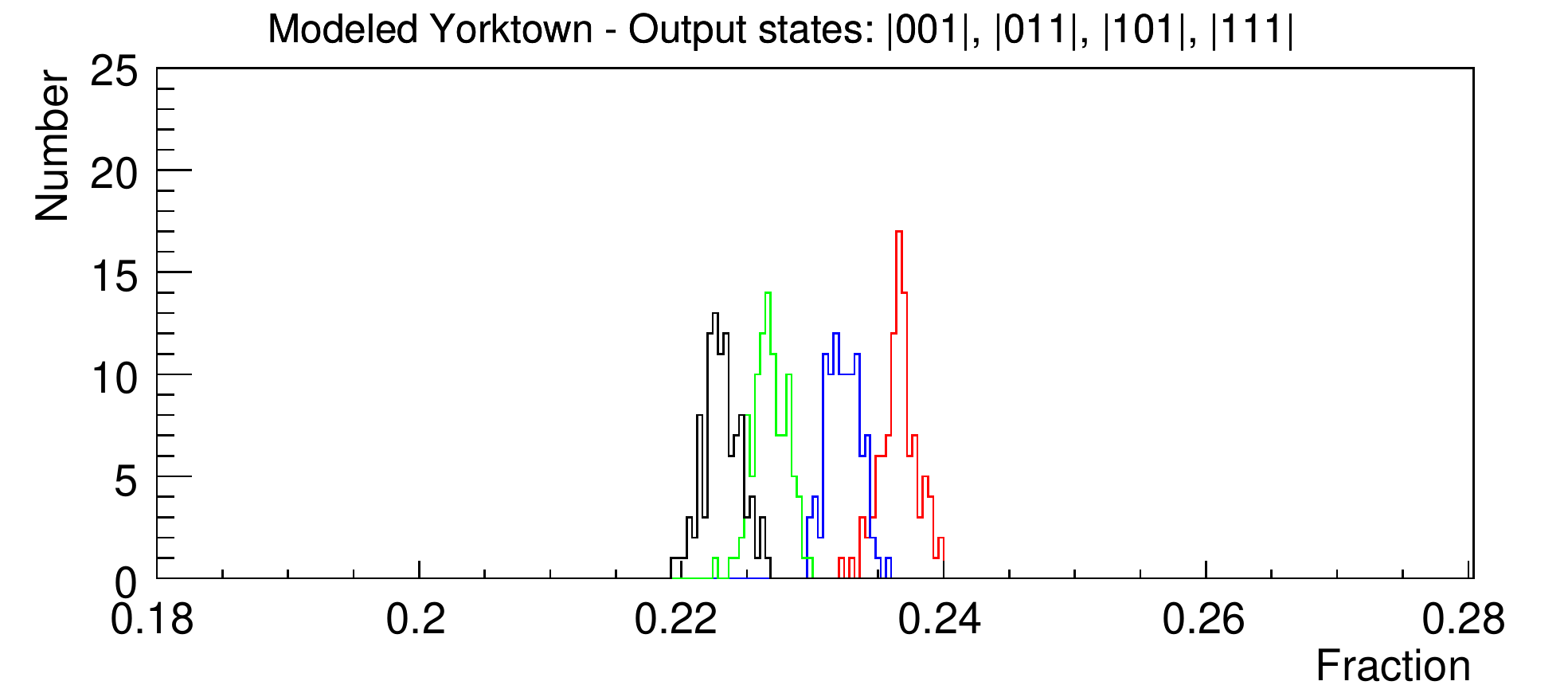}
\end{subfigure}
%%%\vspace{-1.2em}
%%%\end{center}
%\phantom{---} \\
\quad (f)%%%~\vspace{-1.2em}
%%%\begin{center}
\begin{subfigure}{0.95\columnwidth}
  %\centering
  \includegraphics[width=\columnwidth,center]{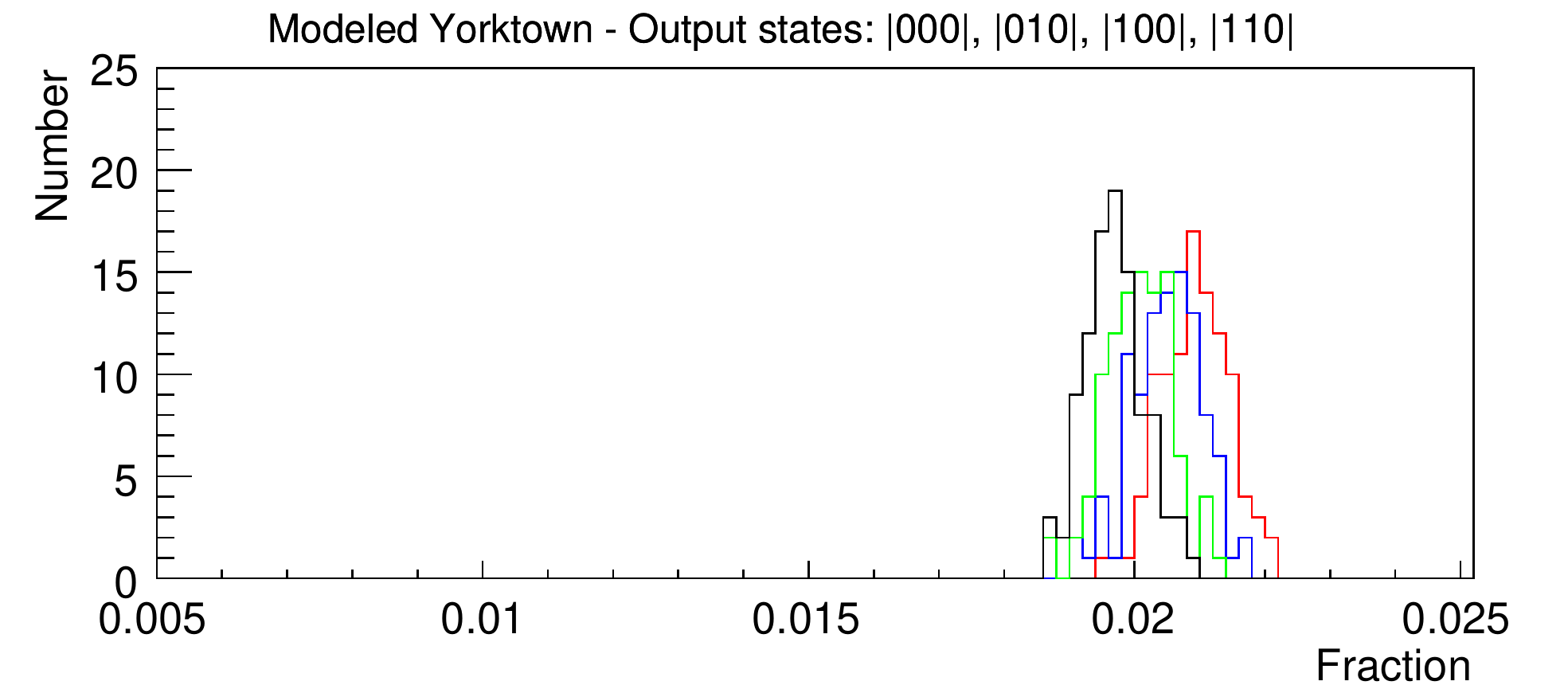}
\end{subfigure}
%%%\vspace{-0.6em}
%%%\end{center}
%%%\caption{CAPTION IF TWO TABLES}
%%%\label{fig:balancedDJ-xx1}
%%%\end{figure}
%
%%%\begin{figure}[htb!]
%%%\raggedright
%%%\begin{center}
%%%    \underline{\textbf{Balanced Deutsch-Jozsa calculation results - II}} \\
%%%    \vspace{-0.6em}
%%%\end{center}
(c)%%%~\vspace{-1.2em}
%%%\begin{center}
\begin{subfigure}{0.95\columnwidth}
  %\centering
  \includegraphics[width=\columnwidth,center]{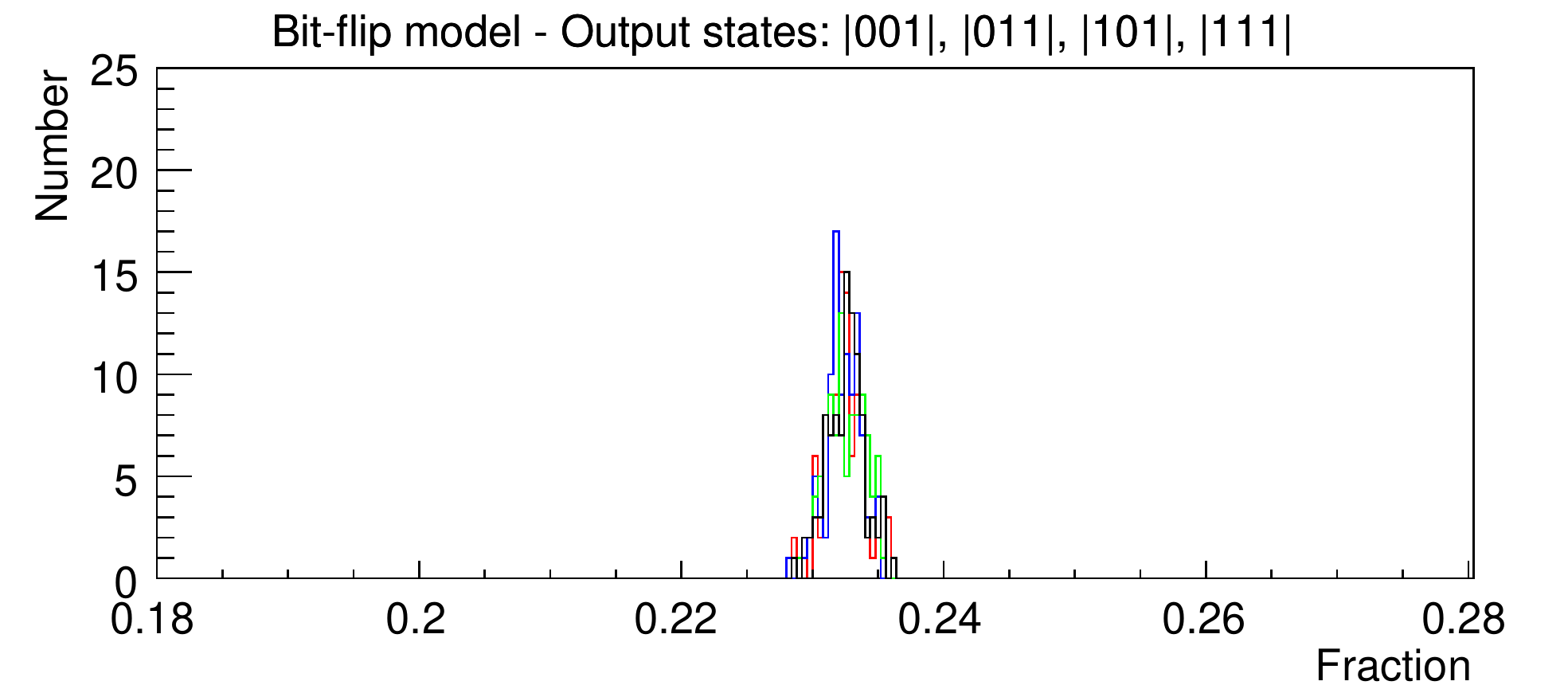}
\end{subfigure}
%%%\vspace{-1.2em}
%%%\end{center}
%\phantom{---} \\
\quad (g)%%%~\vspace{-1.2em}
%%%\begin{center}
\begin{subfigure}{0.95\columnwidth}
  %\centering
  \includegraphics[width=\columnwidth,center]{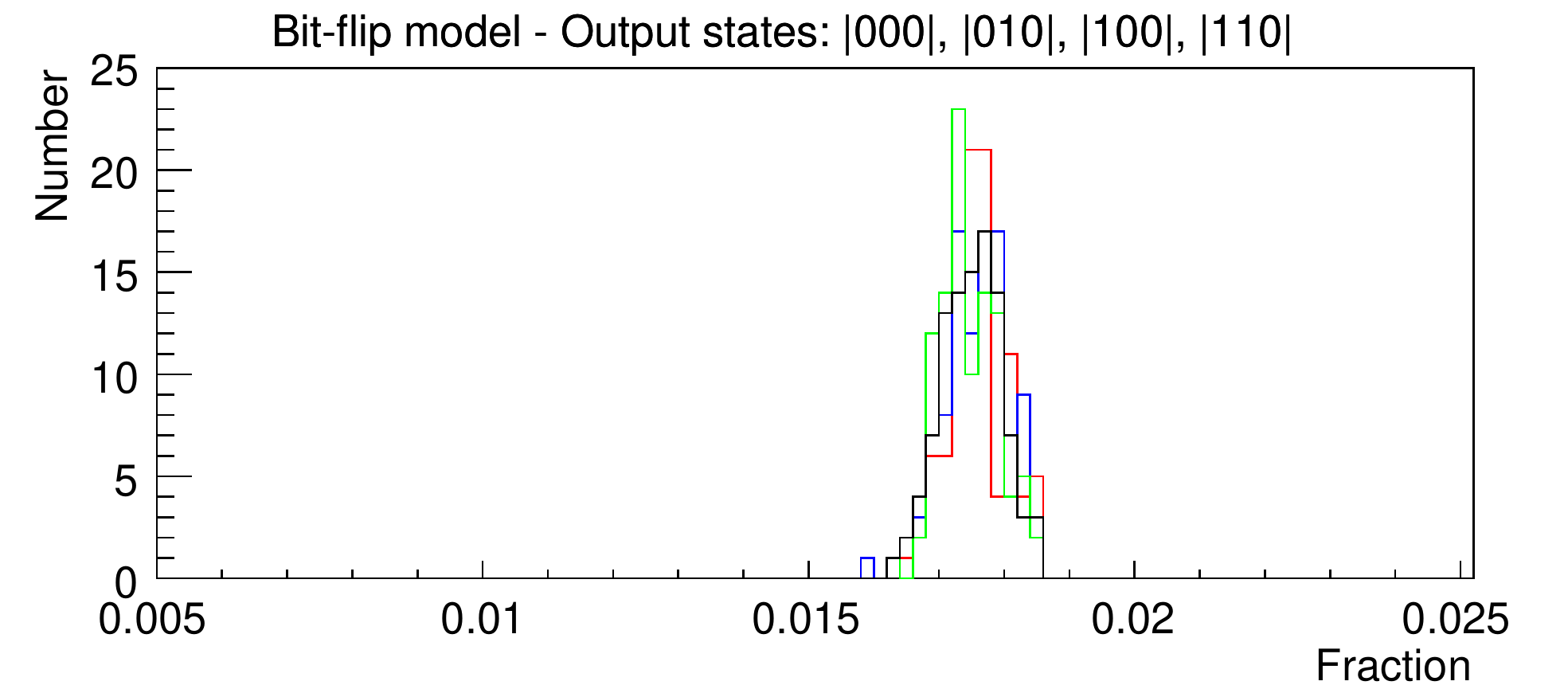}
\end{subfigure}
%%%\vspace{-1.2em}
%%%\end{center}
%\phantom{---} \\
(d)%%%~\vspace{-1.2em}
%%%\begin{center}
\begin{subfigure}{0.95\columnwidth}
  %\centering
  \includegraphics[width=\columnwidth,center]{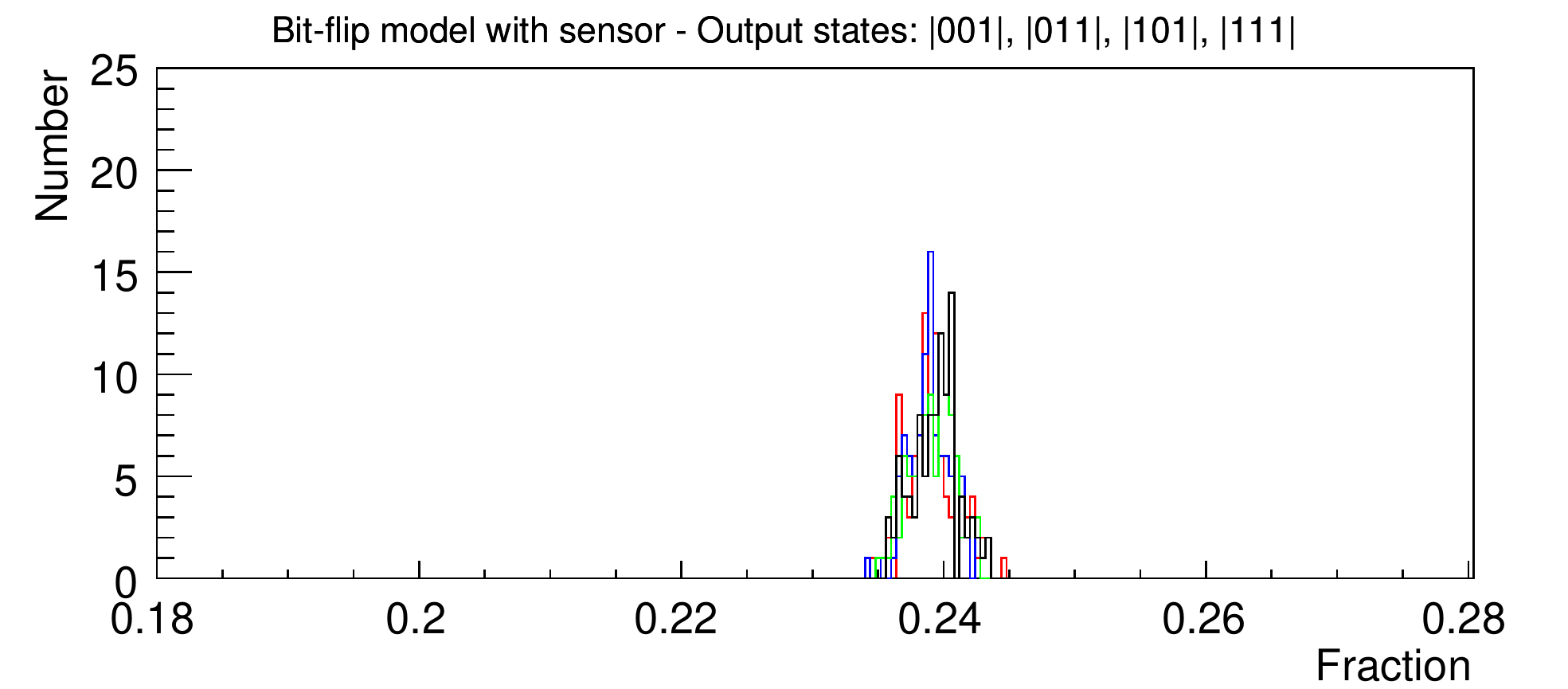}
\end{subfigure}
%%%\end{center}
%\phantom{---} \\
\quad (h)%%%~\vspace{-1.2em}
%%%\begin{center}
\begin{subfigure}{0.95\columnwidth}
  %\centering
  \includegraphics[width=\columnwidth,center]{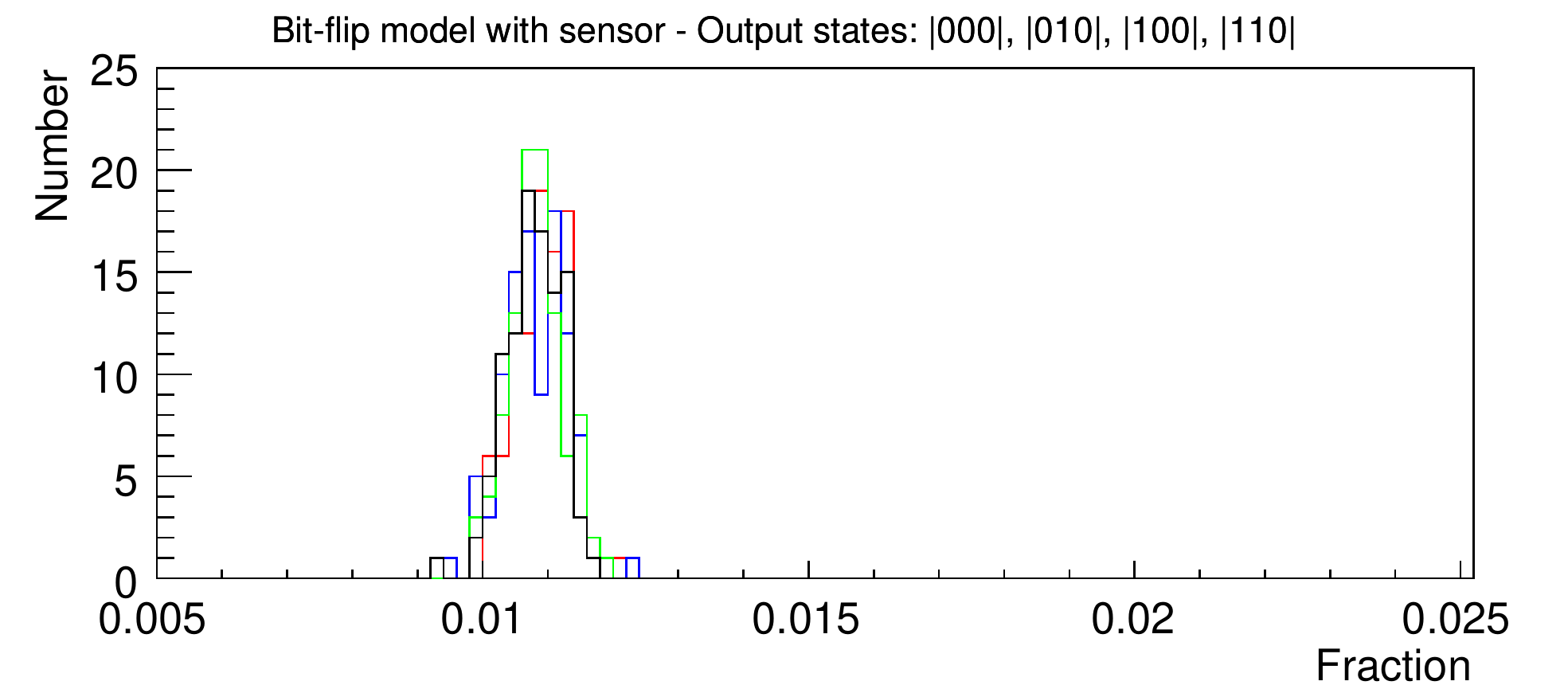}
\end{subfigure}
%%%\vspace{-0.6em}
%%%\end{center}
\vspace{1em}
\caption{Results from three implementations of a balanced Deutsch-Jozsa calculation (see Fig.~\ref{fig:balanced-dj-circuit}). The fraction of results in each of eight outcome cases are shown for the calculations. One hundred repeated sets of 81,920 calculations are displayed. See Figure~\ref{fig:balanced-dj} for an example of the results for a single set. Recall, in an error-free calculation, all four result outcomes in the left column would have 0.25 fractional weighting, each with a Poissonian distribution. Furthermore, in such an error-free case, exactly \emph{zero} outcomes would appear in the right column.}
\label{fig:balancedDJ}
\end{figure*}

Here we present more statistics of the balanced Deutsch-Jozsa calculation presented in the main report. The key interest is related to whether the enhanced performance of the hypothesized sensor-assisted computation result is statistically significant. One hundred trials of 81,920 shots were conducted to determine the variation of the sample. Figure~\ref{fig:balancedDJ} presents the distributions of these one hundred trails of 81,920 shots. The Yorktown backend shows greater variability (Fig.~\ref{fig:balancedDJ}(a)~\&~(e)~), suggesting error inducing effects beyond simple Poisson statistical variation.\footnote{This is also likely a result from the IBM Q Experience's transpilation step for implementation of a quantum circuit on a specific backend as well as errors introduced solely in the measurement stage.} The noise model for the Yorktown backend (Fig.~\ref{fig:balancedDJ}(b)~\&~(f)~), however, shows Poissonian statistical variation, as expected for a fixed, deterministic simulation process. It is interesting to note the modeled noise for the Yorktown backend does not appear to closely match the results of the actual device, and in fact produced ``incorrect'' state outcomes in a larger fraction of calculations (i.e., Fig.~\ref{fig:balancedDJ}(f)~). As expected, the bit-flip-based error models (Fig.~\ref{fig:balancedDJ}(c,d)~\&~(g,h)~) show only Poissonian statistical variation as the errors are discrete in nature and follow a strict fixed probability of being introduced into the quantum circuit by construction. It is clear from these results the improvement provided by the use of the co-located sensor to ``veto'' some calculations does produce a statistically significant enhancement to the computational result when comparing the two bit-flip-based modeled error cases. That is, comparing Figure~\ref{fig:balancedDJ}(d)~to~(c) shows a greater fraction of correct outcome states, while comparing Figure~\ref{fig:balancedDJ}(h)~to~(g) shows a lower fraction of incorrect outcome states.

% Two error system
\section{Probabilities in two error systems}

In this section we analyze, in an entirely generic way, the probability outcomes for two independent, random, bi-modal processes (see Figure~\ref{fig:outcome_table}) on three independent channels. Consider two independent random event processes, each having fixed probabilities, $o$ and $p$, for occurring in a given time period. We refer to these as Type-$o$ and Type-$p$ events in the context of this report. Initially, we make no assumptions about what these events represent. We are interested in detailing all possible ways these two independent events can occur in the given time period. None of the following discussion relies on any quantum mechanical assumptions whatsoever or any knowledge of the event type. There are only four possible cases, as presented in Figure~\ref{fig:outcome_table}.

Since Figure~\ref{fig:outcome_table} is complete and exhaustive of all possibilities, we can write two probability equations to represent the probability of at least one error occurring and the probability of no error occurring, respectively:
\begin{eqnarray}
    P_{\mathrm{error}} & = & o \cdot p + o \cdot (1-p) + (1-o) \cdot p \\
             & = & o + p - o p  \equiv \hat{P}
\end{eqnarray}
\begin{eqnarray}
    P_{\mathrm{no~error}} & = & (1-o) \cdot (1-p) \\
          & = & 1 - p - o + o p \\
          & = & 1 - ( o + p - o p ) = 1 - \hat{P}
\end{eqnarray}
Thus, for two independently drawn random bi-modal errors, the combined probability of an event is $\hat{P}$, while the probability of no event is $(1-\hat{P})$. In this work we will identify the Type-$o$ and Type-$p$ events with different sorts of errors induced on a qubit. We further assume the probability $(1-\hat{P})$ satisfies the quantum error correction requirement of being ``small'' (i.e., less than $0.5$). It should be noted that if $\hat{P}$ is small, then $o$ and $p$ must also both individually be small and therefore also less than $0.5$.

Now consider the outcome equation for three qubits, $\mathrm{q0}$, $\mathrm{q1}$, $\mathrm{q2}$, in one time period when an error may occur on any combination of qubits. We write this as,
\begin{eqnarray}
    \mathbf{1}_{\mathrm{q0,q1,q2}} & = &  \mathbf{1}_{\mathrm{q0}} \times \mathbf{1}_{\mathrm{q1}} \times \mathbf{1}_{\mathrm{q2}} \\
    & = & \big( (1 - \hat{P}) + \hat{P} \big)_{\mathrm{q0}} \\
    &   & \times \big( (1 - \hat{P}) + \hat{P} \big)_{\mathrm{q1}} \\
    &   & \times \big( (1 - \hat{P}) + \hat{P} \big)_{\mathrm{q2}}
\end{eqnarray}
which exhausts all possible outcomes for the three qubits. We further expand this outcome equation to highlight the individual Type-$o$ and Type-$p$ errors, making the compacting notation adjustments, $\bar{o}=(1-o)$ and $\bar{p}=(1-p)$,
\begin{eqnarray}
    \mathbf{1}_{\mathrm{q0,q1,q2}} & = & \big( \bar{o} \bar{p} + o p + o \bar{p} + \bar{o} p \big)_{\mathrm{q0}} \\
    &   & \times \big( \bar{o} \bar{p} + o p + o \bar{p} + \bar{o} p  \big)_{\mathrm{q1}} \\
    &   & \times \big( \bar{o} \bar{p} + o p + o \bar{p} + \bar{o} p  \big)_{\mathrm{q2}}
\end{eqnarray}

Given the assumption in this report that all error types are bit-flip errors, the terms $o p$ have special significance in that the two errors on a single qubit will cancel out. Thus, we add a notation, $\bar{c}$, representing when errors cancel out:
\begin{eqnarray}
    \mathbf{1}_{\mathrm{q0,q1,q2}} & = & \big( \bar{o} \bar{p} + \bar{c} + o \bar{p} + \bar{o} p \big)_{\mathrm{q0}} \\
    &   & \times \big( \bar{o} \bar{p} + \bar{c} + o \bar{p} + \bar{o} p  \big)_{\mathrm{q1}} \\
    &   & \times \big( \bar{o} \bar{p} + \bar{c} + o \bar{p} + \bar{o} p  \big)_{\mathrm{q2}}
\end{eqnarray}

At this point we assume the Type-$o$ and Type-$p$ errors, while independent between the three qubits, come from the same physical source types and have the same probability values (i.e., $o = o_{\mathrm{q0}}=o_{\mathrm{q1}}=o_{\mathrm{q2}}$ and $p = p_{\mathrm{q0}}=p_{\mathrm{q1}}=p_{\mathrm{q2}}$). Thus, multiplying through the outcome equation and regrouping terms associated with order of $\bar{o}$-factors, we arrive at:
\begin{eqnarray}
    \mathbf{1}_{\mathrm{q0,q1,q2}} & = & \bar{o}^3 \left(3 p \bar{p}^2+\bar{p}^3\right) \\
    &   & +\ \bar{o}^3 \left(p^3+3 p^2 \bar{p}\right) \\
    &   & +\ \bar{o}^2 \left(6 p \bar{c} \bar{p}+3 \bar{c} \bar{p}^2+3 o \bar{p}^3\right) \\
    &   & +\ \bar{o}^2 \left(3 p^2 \bar{c}+3 o p^2 \bar{p}+6 o p \bar{p}^2\right) \\
    &   & +\ \bar{o} \left(3 p \bar{c}^2+3 \bar{c}^2 \bar{p}+6 o \bar{c} \bar{p}^2\right) \\
    &   & +\ \bar{o} \left(6 o p \bar{c} \bar{p}+3 o^2 p \bar{p}^2+3 o^2 \bar{p}^3\right) \\
    &   & +\ \bar{c}^3+3 o \bar{c}^2 \bar{p} \\
    &   & +\ 3 o^2 \bar{c} \bar{p}^2+o^3 \bar{p}^3
\end{eqnarray}

We identify Type-$o$ errors with environmental, sensor-detectable errors and Type-$p$ errors with entanglement type errors, which cannot be detected. We regroup the terms related to whether the errors are correctable (C), faulty (F), or correctable via cancellation (CC) as well as the distinction of whether the sensor-assist either outright REJECTs the calculation (R$_{\mathrm{S}}$) or sets the REJECT flag based on the syndrome parity test (R$_{\mathrm{PT}}$). These amount to the fractions, $\mathcal{F}$, of cases of each kind.
\begin{eqnarray}
    \mathcal{F}_{\mathrm{C\textit{vs.}C}} & = & \bar{o}^3 \left(3 p \bar{p}^2+\bar{p}^3\right) + \bar{o}^2 \left(3 o \bar{p}^3\right) \\
    \mathcal{F}_{\mathrm{CC\textit{vs.}CC}} & = & \bar{o}^2 \left(3 \bar{c} \bar{p}^2\right) \\
    \mathcal{F}_{\mathrm{F\textit{vs.}F}} & = & \bar{o}^3 \left(p^3+3 p^2 \bar{p}\right) \\
       &   & +\ \bar{o}^2 \left(3 p^2 \bar{c}+3 o p^2 \bar{p}\right) \\
    \mathcal{F}_{\mathrm{CC\textit{vs.}R}_{\mathrm{PT}}} & = & \bar{o}^2 \left(6 p \bar{c} \bar{p}\right) \\
    \mathcal{F}_{\mathrm{F\textit{vs.}R}_{\mathrm{PT}}} & = & \bar{o}^2 \left(6 o p \bar{p}^2\right) \\ 
    \mathcal{F}_{\mathrm{CC\textit{vs.}R}_{\mathrm{S}}} & = & \bar{o} \left(3 p \bar{c}^2+3 \bar{c}^2 \bar{p}+6 o \bar{c} \bar{p}^2\right) \\
       &   & +\ \bar{c}^3+3 o \bar{c}^2 \bar{p} \\
    \mathcal{F}_{\mathrm{F\textit{vs.}R}_{\mathrm{S}}} & = & \bar{o} \left(6 o p \bar{c} \bar{p}+3 o^2 p \bar{p}^2+3 o^2 \bar{p}^3\right) \\ 
       &   & +\ 3 o^2 \bar{c} \bar{p}^2+o^3 \bar{p}^3
\end{eqnarray}

\begin{figure}[ht!]
    \centering
    \includegraphics[width=\columnwidth]{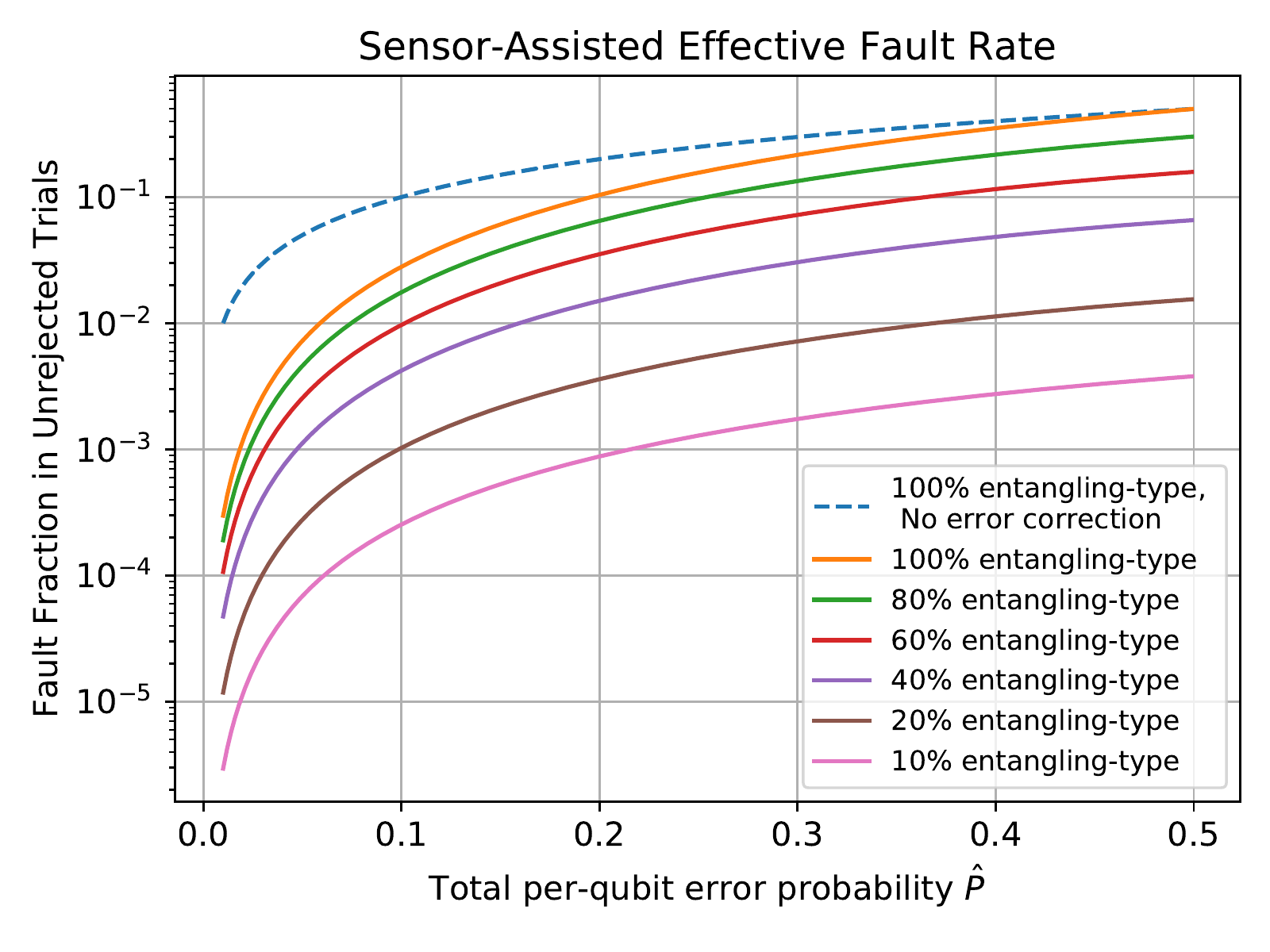}
    \caption{Fraction of calculations containing faults that are not rejected by combined information of the co-sensors and parity registers, as a function of the total per-qubit error rate $\hat{P}$, and the fraction of errors that are of an entangling type (Type-$p$), which are not sensor-detectable.  In terms of the final state truth table, this is defined as $ \mathcal{F}_{\mathrm{F\textit{vs.}F}} / ( \mathcal{F}_{\mathrm{F\textit{vs.}F}} + \mathcal{F}_{\mathrm{C\textit{vs.}C}} + \mathcal{F}_{\mathrm{CC\textit{vs.}CC}} ) $.}
    \label{fig:eff_fault}
\end{figure}

An alternative means for presenting the outcomes is through the truth table of all 64 possible error combinations. This is presented in Tables~\ref{tab:all_64_cases_I}~\&~\ref{tab:all_64_cases_II}. For specific numerical cases, see Figure~\ref{fig:eff_fault}, which shows the effective fault rate (fraction of faults in unrejected calculations, i.e., $ \mathcal{F}_{\mathrm{F\textit{vs.}F}} / ( \mathcal{F}_{\mathrm{F\textit{vs.}F}} + \mathcal{F}_{\mathrm{C\textit{vs.}C}} + \mathcal{F}_{\mathrm{CC\textit{vs.}CC}} ) $ calculated from equations C24-C27).

\def\arraystretch{1.1}

\begin{table*}[h]
    \small
    \centering
    \begin{tabular}{|cc|ccc|c|c|c|}
    \hline
\multicolumn{2}{|c|}{\textbf{Errors}} & \multicolumn{3}{c|}{\textbf{Gates}} & \textbf{Synd.} & \textbf{Prob.} & \textbf{Outcome} \\
\multicolumn{2}{|c|}{[Enviro.]} & \multicolumn{3}{c|}{Col. 5~\&~6} & Ancilla & Error $\times$ & Standard \\
\multicolumn{2}{|c|}{(Entangle)} & \multicolumn{3}{c|}{$\Rightarrow$ Result} & c-reg. & Non-error &  \textit{vs.} Assisted  \\
    \hline \hline
~[000]  & (000) & ~\textbf{\texttt{II}}  &      & \textbf{\texttt{I}} &  &  $o^{0} \cdot p^{0}$ &  \phantom{AC  \textit{vs.}  R$_{\mathrm{PT}}$} \\
         &         & ~\textbf{\texttt{II}}  & ~$\Rightarrow$  & \textbf{\texttt{I}} & 0\texttt{x}0 & $\times$ &  C  \textit{vs.}  C \\
         &         & ~\textbf{\texttt{II}}  &      & \textbf{\texttt{I}} &  &  $\bar{o}^{3} \cdot \bar{p}^{3}$ &   \\ \hline
         & (001) & ~\textbf{\texttt{IX}}  &      & \textbf{\texttt{X}} &  &  $o^{0} \cdot p^{1}$ &   \\
         &         & ~\textbf{\texttt{II}}  & ~$\Rightarrow$  & \textbf{\texttt{I}} & 0\texttt{x}3 & $\times$ & C  \textit{vs.}  C  \\
         &         & ~\textbf{\texttt{II}}  &      & \textbf{\texttt{I}} &  &  $\bar{o}^{3} \cdot \bar{p}^{2}$ &    \\ \hline
         & (010) & ~\textbf{\texttt{II}}  &      & \textbf{\texttt{I}} &  &  $o^{0} \cdot p^{1}$  &   \\
         &         & ~\textbf{\texttt{IX}}  & ~$\Rightarrow$  & \textbf{\texttt{X}} & 0\texttt{x}1 & $\times$ & C  \textit{vs.}  C  \\
         &         & ~\textbf{\texttt{II}}  &      & \textbf{\texttt{I}} &  &  $\bar{o}^{3} \cdot \bar{p}^{2}$ &    \\ \hline
         & (100) & ~\textbf{\texttt{II}}  &      & \textbf{\texttt{I}} &  &  $o^{0} \cdot p^{1}$  &   \\
         &         & ~\textbf{\texttt{II}}  & ~$\Rightarrow$  & \textbf{\texttt{I}} & 0\texttt{x}2 & $\times$ & C  \textit{vs.}  C  \\
         &         & ~\textbf{\texttt{IX}}  &      & \textbf{\texttt{X}} &  &  $\bar{o}^{3} \cdot \bar{p}^{2}$ &    \\ \hline
         & (011) & ~\textbf{\texttt{IX}}  &      & \textbf{\texttt{X}} &  &  $o^{0} \cdot p^{2}$  &   \\
         &         & ~\textbf{\texttt{IX}}  & ~$\Rightarrow$  & \textbf{\texttt{X}} & 0\texttt{x}2 & $\times$ & F  \textit{vs.}  F  \\
         &         & ~\textbf{\texttt{II}}  &      & \textbf{\texttt{I}} &  &  $\bar{o}^{3} \cdot \bar{p}^{1}$ &    \\ \hline
         & (101) & ~\textbf{\texttt{IX}}  &      & \textbf{\texttt{X}} &  &  $o^{0} \cdot p^{2}$  &   \\
         &         & ~\textbf{\texttt{II}}  & ~$\Rightarrow$  & \textbf{\texttt{I}} & 0\texttt{x}1 & $\times$ & F  \textit{vs.}  F  \\
         &         & ~\textbf{\texttt{IX}}  &      & \textbf{\texttt{X}} &  &  $\bar{o}^{3} \cdot \bar{p}^{1}$ &    \\ \hline
         & (110) & ~\textbf{\texttt{II}}  &      & \textbf{\texttt{I}} &  &  $o^{0} \cdot p^{2}$  &   \\
         &         & ~\textbf{\texttt{IX}}  & ~$\Rightarrow$  & \textbf{\texttt{X}} & 0\texttt{x}3 & $\times$ & F  \textit{vs.}  F  \\
         &         & ~\textbf{\texttt{IX}}  &      & \textbf{\texttt{X}} &  &  $\bar{o}^{3} \cdot \bar{p}^{1}$ &    \\ \hline
         & (111) & ~\textbf{\texttt{IX}}  &      & \textbf{\texttt{X}} &  &  $o^{0} \cdot p^{3}$  &   \\
         &         & ~\textbf{\texttt{IX}}  & ~$\Rightarrow$  & \textbf{\texttt{X}} & 0\texttt{x}0 & $\times$ & F  \textit{vs.}  F  \\
         &         & ~\textbf{\texttt{IX}}  &      & \textbf{\texttt{X}} &  &  $\bar{o}^{3} \cdot \bar{p}^{0}$ &    \\ \hline
    \end{tabular}
    %\caption{Table of.}
    %\label{tab:all_64_cases}
%\end{table}
%\phantom{-|-}
%\begin{table}[h]
    \small
    \centering
    \begin{tabular}{|cc|ccc|c|c|c|}
    \hline
\multicolumn{2}{|c|}{\textbf{Errors}} & \multicolumn{3}{c|}{\textbf{Gates}} & \textbf{Synd.} & \textbf{Prob.} & \textbf{Outcome} \\
\multicolumn{2}{|c|}{[Enviro.]} & \multicolumn{3}{c|}{Col. 5~\&~6} & Ancilla & Error $\times$ & Standard \\
\multicolumn{2}{|c|}{(Entangle)} & \multicolumn{3}{c|}{$\Rightarrow$ Result} & c-reg. & Non-error &  \textit{vs.} Assisted  \\
    \hline \hline
~[001]  & (000) & ~\textbf{\texttt{XI}}  &      & \textbf{\texttt{X}} &  &  $o^{1} \cdot p^{0}$ & \phantom{CC  \textit{vs.}  R$_{\mathrm{PT}}$}  \\
         &         & ~\textbf{\texttt{II}}  & ~$\Rightarrow$  & \textbf{\texttt{I}} & 0\texttt{x}3 & $\times$ &  C  \textit{vs.}  C \\
         &         & ~\textbf{\texttt{II}}  &      & \textbf{\texttt{I}} &  &  $\bar{o}^{2} \cdot \bar{p}^{3}$ &    \\ \hline
         & (001) & ~\textbf{\texttt{XX}}  &      & \textbf{\texttt{I}} &  &  $o^{1} \cdot p^{1}$ &   \\
         &         & ~\textbf{\texttt{II}}  & ~$\Rightarrow$  & \textbf{\texttt{I}} & 0\texttt{x}0 & $\times$ & CC  \textit{vs.}  CC  \\
         &         & ~\textbf{\texttt{II}}  &      & \textbf{\texttt{I}} &  &  $\bar{o}^{2} \cdot \bar{p}^{2}$ &    \\ \hline
         & (010) & ~\textbf{\texttt{XI}}  &      & \textbf{\texttt{X}} &  &  $o^{1} \cdot p^{1}$  &   \\
         &         & ~\textbf{\texttt{IX}}  & ~$\Rightarrow$  & \textbf{\texttt{X}} & 0\texttt{x}2 & $\times$ &  F  \textit{vs.}  R$_{\mathrm{PT}}$ \\
         &         & ~\textbf{\texttt{II}}  &      & \textbf{\texttt{I}} &  &  $\bar{o}^{2} \cdot \bar{p}^{2}$ &    \\ \hline
         & (100) & ~\textbf{\texttt{XI}}  &      & \textbf{\texttt{X}} &  &  $o^{1} \cdot p^{1}$  &   \\
         &         & ~\textbf{\texttt{II}}  & ~$\Rightarrow$  & \textbf{\texttt{I}} & 0\texttt{x}1 & $\times$ & F  \textit{vs.}  R$_{\mathrm{PT}}$  \\
         &         & ~\textbf{\texttt{IX}}  &      & \textbf{\texttt{X}} &  &  $\bar{o}^{2} \cdot \bar{p}^{2}$ &    \\ \hline
         & (011) & ~\textbf{\texttt{XX}}  &      & \textbf{\texttt{I}} &  &  $o^{1} \cdot p^{2}$ &   \\
         &         & ~\textbf{\texttt{IX}}  & ~$\Rightarrow$  & \textbf{\texttt{X}} & 0\texttt{x}1 & $\times$ & CC  \textit{vs.}  R$_{\mathrm{PT}}$  \\
         &         & ~\textbf{\texttt{II}}  &      & \textbf{\texttt{I}} &  &  $\bar{o}^{2} \cdot \bar{p}^{1}$ &    \\ \hline
         & (101) & ~\textbf{\texttt{XX}}  &      & \textbf{\texttt{I}} &  &  $o^{1} \cdot p^{2}$  &   \\
         &         & ~\textbf{\texttt{II}}  & ~$\Rightarrow$  & \textbf{\texttt{I}} & 0\texttt{x}2 & $\times$ &  CC  \textit{vs.}  R$_{\mathrm{PT}}$ \\
         &         & ~\textbf{\texttt{IX}}  &      & \textbf{\texttt{X}} &  &  $\bar{o}^{2} \cdot \bar{p}^{1}$ &    \\ \hline
         & (110) & ~\textbf{\texttt{XI}}  &      & \textbf{\texttt{X}} &  &  $o^{1} \cdot p^{2}$  &   \\
         &         & ~\textbf{\texttt{IX}}  & ~$\Rightarrow$  & \textbf{\texttt{X}} & 0\texttt{x}0 & $\times$ & F  \textit{vs.}  F  \\
         &         & ~\textbf{\texttt{IX}}  &      & \textbf{\texttt{X}} &  &  $\bar{o}^{2} \cdot \bar{p}^{1}$ &    \\ \hline
         & (111) & ~\textbf{\texttt{XX}}  &      & \textbf{\texttt{I}} &  &  $o^{1} \cdot p^{3}$  &   \\
         &         & ~\textbf{\texttt{IX}}  & ~$\Rightarrow$  & \textbf{\texttt{X}} & 0\texttt{x}3 & $\times$ &  F  \textit{vs.}  F \\
         &         & ~\textbf{\texttt{IX}}  &      & \textbf{\texttt{X}} &  &  $\bar{o}^{2} \cdot \bar{p}^{0}$ &    \\ \hline
    \end{tabular}
    %\caption{Table of.}
    %\label{tab:all_64_cases}
%\end{table}
%\phantom{-|-}
%\begin{table}[h]
    \small
    \centering
    \begin{tabular}{|cc|ccc|c|c|c|}
    \hline
\multicolumn{2}{|c|}{\textbf{Errors}} & \multicolumn{3}{c|}{\textbf{Gates}} & \textbf{Synd.} & \textbf{Prob.} & \textbf{Outcome} \\
\multicolumn{2}{|c|}{[Enviro.]} & \multicolumn{3}{c|}{Col. 5~\&~6} & Ancilla & Error $\times$ & Standard \\
\multicolumn{2}{|c|}{(Entangle)} & \multicolumn{3}{c|}{$\Rightarrow$ Result} & c-reg. & Non-error &  \textit{vs.} Assisted  \\
    \hline \hline
~[010]  & (000) & ~\textbf{\texttt{II}}  &      & \textbf{\texttt{I}} &  &  $o^{1} \cdot p^{0}$  &  \phantom{CC  \textit{vs.}  R$_{\mathrm{PT}}$} \\
         &         & ~\textbf{\texttt{XI}}  & ~$\Rightarrow$  & \textbf{\texttt{X}} & 0\texttt{x}1 & $\times$ & C  \textit{vs.}  C  \\
         &         & ~\textbf{\texttt{II}}  &      & \textbf{\texttt{I}} &  &  $\bar{o}^{2} \cdot \bar{p}^{3}$ &    \\ \hline
         & (001) & ~\textbf{\texttt{IX}}  &      & \textbf{\texttt{X}} &  &  $o^{1} \cdot p^{1}$  &   \\
         &         & ~\textbf{\texttt{XI}}  & ~$\Rightarrow$  & \textbf{\texttt{X}} & 0\texttt{x}2 & $\times$ & F  \textit{vs.}  R$_{\mathrm{PT}}$  \\
         &         & ~\textbf{\texttt{II}}  &      & \textbf{\texttt{I}} &  &  $\bar{o}^{2} \cdot \bar{p}^{2}$ &    \\ \hline
         & (010) & ~\textbf{\texttt{II}}  &      & \textbf{\texttt{I}} &  &  $o^{1} \cdot p^{1}$  &   \\
         &         & ~\textbf{\texttt{XX}}  & ~$\Rightarrow$  & \textbf{\texttt{I}} & 0\texttt{x}0 & $\times$ & CC  \textit{vs.}  CC  \\
         &         & ~\textbf{\texttt{II}}  &      & \textbf{\texttt{I}} &  &  $\bar{o}^{2} \cdot \bar{p}^{2}$ &    \\ \hline
         & (100) & ~\textbf{\texttt{II}}  &      & \textbf{\texttt{I}} &  &  $o^{1} \cdot p^{1}$   &   \\
         &         & ~\textbf{\texttt{XI}}  & ~$\Rightarrow$  & \textbf{\texttt{X}} & 0\texttt{x}3 & $\times$ & F  \textit{vs.}  R$_{\mathrm{PT}}$  \\
         &         & ~\textbf{\texttt{IX}}  &      & \textbf{\texttt{X}} &  &  $\bar{o}^{2} \cdot \bar{p}^{2}$ &    \\ \hline
         & (011) & ~\textbf{\texttt{IX}}  &      & \textbf{\texttt{X}} &  &  $o^{1} \cdot p^{2}$  &   \\
         &         & ~\textbf{\texttt{XX}}  & ~$\Rightarrow$  & \textbf{\texttt{I}} & 0\texttt{x}3 & $\times$ & CC  \textit{vs.}  R$_{\mathrm{PT}}$  \\
         &         & ~\textbf{\texttt{II}}  &      & \textbf{\texttt{I}} &  &  $\bar{o}^{2} \cdot \bar{p}^{1}$ &    \\ \hline
         & (101) & ~\textbf{\texttt{IX}}  &      & \textbf{\texttt{X}} &  &  $o^{1} \cdot p^{2}$   &   \\
         &         & ~\textbf{\texttt{XI}}  & ~$\Rightarrow$  & \textbf{\texttt{X}} & 0\texttt{x}0 & $\times$ & F  \textit{vs.}  F  \\
         &         & ~\textbf{\texttt{IX}}  &      & \textbf{\texttt{X}} &  &  $\bar{o}^{2} \cdot \bar{p}^{1}$ &    \\ \hline
         & (110) & ~\textbf{\texttt{II}}  &      & \textbf{\texttt{I}} &  &  $o^{1} \cdot p^{2}$  &   \\
         &         & ~\textbf{\texttt{XX}}  & ~$\Rightarrow$  & \textbf{\texttt{I}} & 0\texttt{x}2 & $\times$ & CC  \textit{vs.}  R$_{\mathrm{PT}}$  \\
         &         & ~\textbf{\texttt{IX}}  &      & \textbf{\texttt{X}} &  &  $\bar{o}^{2} \cdot \bar{p}^{1}$ &    \\ \hline
         & (111) & ~\textbf{\texttt{IX}}  &      & \textbf{\texttt{X}} &  &  $o^{1} \cdot p^{3}$  &   \\
         &         & ~\textbf{\texttt{XX}}  & ~$\Rightarrow$  & \textbf{\texttt{I}} & 0\texttt{x}1 & $\times$ &  F  \textit{vs.}  F \\
         &         & ~\textbf{\texttt{IX}}  &      & \textbf{\texttt{X}} &  &  $\bar{o}^{2} \cdot \bar{p}^{0}$ &    \\ \hline
    \end{tabular}
    %\caption{Table of.}
    %\label{tab:all_64_cases}
%\end{table}
%\phantom{-|-}
%\begin{table}[h]
    \small
    \centering
    \begin{tabular}{|cc|ccc|c|c|c|}
    \hline
\multicolumn{2}{|c|}{\textbf{Errors}} & \multicolumn{3}{c|}{\textbf{Gates}} & \textbf{Synd.} & \textbf{Prob.} & \textbf{Outcome} \\
\multicolumn{2}{|c|}{[Enviro.]} & \multicolumn{3}{c|}{Col. 5~\&~6} & Ancilla & Error $\times$ & Standard \\
\multicolumn{2}{|c|}{(Entangle)} & \multicolumn{3}{c|}{$\Rightarrow$ Result} & c-reg. & Non-error &  \textit{vs.} Assisted  \\
    \hline \hline
~[100]  & (000) & ~\textbf{\texttt{II}}  &      & \textbf{\texttt{I}} &  &  $o^{1} \cdot p^{0}$  & \phantom{CC  \textit{vs.}  R$_{\mathrm{PT}}$}  \\
         &         & ~\textbf{\texttt{II}}  & ~$\Rightarrow$  & \textbf{\texttt{I}} & 0\texttt{x}2 & $\times$ & C  \textit{vs.}  C  \\
         &         & ~\textbf{\texttt{XI}}  &      & \textbf{\texttt{X}} &  &  $\bar{o}^{2} \cdot \bar{p}^{3}$ &    \\ \hline
         & (001) & ~\textbf{\texttt{IX}}  &      & \textbf{\texttt{X}} &  &  $o^{1} \cdot p^{1}$  &   \\
         &         & ~\textbf{\texttt{II}}  & ~$\Rightarrow$  & \textbf{\texttt{I}} & 0\texttt{x}1 & $\times$ & F  \textit{vs.}  R$_{\mathrm{PT}}$  \\
         &         & ~\textbf{\texttt{XI}}  &      & \textbf{\texttt{X}} &  &  $\bar{o}^{2} \cdot \bar{p}^{2}$ &    \\ \hline
         & (010) & ~\textbf{\texttt{II}}  &      & \textbf{\texttt{I}} &  &  $o^{1} \cdot p^{1}$  &   \\
         &         & ~\textbf{\texttt{IX}}  & ~$\Rightarrow$  & \textbf{\texttt{X}} & 0\texttt{x}3 & $\times$ & F  \textit{vs.}  R$_{\mathrm{PT}}$  \\
         &         & ~\textbf{\texttt{XI}}  &      & \textbf{\texttt{X}} &  &  $\bar{o}^{2} \cdot \bar{p}^{2}$ &    \\ \hline
         & (100) & ~\textbf{\texttt{II}}  &      & \textbf{\texttt{I}} &  &  $o^{1} \cdot p^{1}$  &   \\
         &         & ~\textbf{\texttt{II}}  & ~$\Rightarrow$  & \textbf{\texttt{I}} & 0\texttt{x}0 & $\times$ & CC  \textit{vs.}  CC  \\
         &         & ~\textbf{\texttt{XX}}  &      & \textbf{\texttt{I}} &  &  $\bar{o}^{2} \cdot \bar{p}^{2}$ &    \\ \hline
         & (011) & ~\textbf{\texttt{IX}}  &      & \textbf{\texttt{X}} &  &  $o^{1} \cdot p^{2}$  &   \\
         &         & ~\textbf{\texttt{IX}}  & ~$\Rightarrow$  & \textbf{\texttt{X}} & 0\texttt{x}0 & $\times$ & F  \textit{vs.}  F  \\
         &         & ~\textbf{\texttt{XI}}  &      & \textbf{\texttt{X}} &  &  $\bar{o}^{2} \cdot \bar{p}^{1}$ &    \\ \hline
         & (101) & ~\textbf{\texttt{IX}}  &      & \textbf{\texttt{X}} &  &  $o^{1} \cdot p^{2}$   &   \\
         &         & ~\textbf{\texttt{II}}  & ~$\Rightarrow$  & \textbf{\texttt{I}} & 0\texttt{x}3 & $\times$ & CC  \textit{vs.}  R$_{\mathrm{PT}}$  \\
         &         & ~\textbf{\texttt{XX}}  &      & \textbf{\texttt{I}} &  &  $\bar{o}^{2} \cdot \bar{p}^{1}$ &    \\ \hline
         & (110) & ~\textbf{\texttt{II}}  &      & \textbf{\texttt{I}} &  &  $o^{1} \cdot p^{2}$  &   \\
         &         & ~\textbf{\texttt{IX}}  & ~$\Rightarrow$  & \textbf{\texttt{X}} & 0\texttt{x}1 & $\times$ & CC  \textit{vs.}  R$_{\mathrm{PT}}$  \\
         &         & ~\textbf{\texttt{XX}}  &      & \textbf{\texttt{I}} &  &  $\bar{o}^{2} \cdot \bar{p}^{1}$ &    \\ \hline
         & (111) & ~\textbf{\texttt{IX}}  &      & \textbf{\texttt{X}} &  &  $o^{1} \cdot p^{3}$  &   \\
         &         & ~\textbf{\texttt{IX}}  & ~$\Rightarrow$  & \textbf{\texttt{X}} & 0\texttt{x}2 & $\times$ & F  \textit{vs.}  F  \\
         &         & ~\textbf{\texttt{XX}}  &      & \textbf{\texttt{I}} &  &  $\bar{o}^{2} \cdot \bar{p}^{0}$ &    \\ \hline
    \end{tabular}
    \caption{See main report and Table~\ref{tab:001_cases} for description of tables.}
    \label{tab:all_64_cases_I}
\end{table*}
%\phantom{-|-}
\begin{table*}[h]
    \small
    \centering
    \begin{tabular}{|cc|ccc|c|c|c|}
    \hline
\multicolumn{2}{|c|}{\textbf{Errors}} & \multicolumn{3}{c|}{\textbf{Gates}} & \textbf{Synd.} & \textbf{Prob.} & \textbf{Outcome} \\
\multicolumn{2}{|c|}{[Enviro.]} & \multicolumn{3}{c|}{Col. 5~\&~6} & Ancilla & Error $\times$ & Standard \\
\multicolumn{2}{|c|}{(Entangle)} & \multicolumn{3}{c|}{$\Rightarrow$ Result} & c-reg. & Non-error &  \textit{vs.} Assisted  \\
    \hline \hline
~[011]  & (000) & ~\textbf{\texttt{XI}}  &      & \textbf{\texttt{X}} &  &  $o^{2} \cdot p^{0}$   &  \phantom{CC  \textit{vs.}  R$_{\mathrm{PT}}$} \\
         &         & ~\textbf{\texttt{XI}}  & ~$\Rightarrow$  & \textbf{\texttt{X}} & 0\texttt{x}2 & $\times$ & F  \textit{vs.}  R$_{\mathrm{S}}$  \\
         &         & ~\textbf{\texttt{II}}  &      & \textbf{\texttt{I}} &  &  $\bar{o}^{1} \cdot \bar{p}^{3}$ &    \\ \hline
         & (001) & ~\textbf{\texttt{XX}}  &      & \textbf{\texttt{I}} &  &  $o^{2} \cdot p^{1}$  &   \\
         &         & ~\textbf{\texttt{XI}}  & ~$\Rightarrow$  & \textbf{\texttt{X}} & 0\texttt{x}1 & $\times$ & CC  \textit{vs.}  R$_{\mathrm{S}}$  \\
         &         & ~\textbf{\texttt{II}}  &      & \textbf{\texttt{I}} &  &  $\bar{o}^{1} \cdot \bar{p}^{2}$ &    \\ \hline
         & (010) & ~\textbf{\texttt{XI}}  &      & \textbf{\texttt{X}} &  &  $o^{2} \cdot p^{1}$   &   \\
         &         & ~\textbf{\texttt{XX}}  & ~$\Rightarrow$  & \textbf{\texttt{I}} & 0\texttt{x}3 & $\times$ & CC  \textit{vs.}  R$_{\mathrm{S}}$  \\
         &         & ~\textbf{\texttt{II}}  &      & \textbf{\texttt{I}} &  &  $\bar{o}^{1} \cdot \bar{p}^{2}$ &    \\ \hline
         & (100) & ~\textbf{\texttt{XI}}  &      & \textbf{\texttt{X}} &  &  $o^{2} \cdot p^{1}$   &   \\
         &         & ~\textbf{\texttt{XI}}  & ~$\Rightarrow$  & \textbf{\texttt{X}} & 0\texttt{x}0 & $\times$ & F  \textit{vs.}  R$_{\mathrm{S}}$  \\
         &         & ~\textbf{\texttt{IX}}  &      & \textbf{\texttt{X}} &  &  $\bar{o}^{1} \cdot \bar{p}^{2}$ &    \\ \hline
         & (011) & ~\textbf{\texttt{XX}}  &      & \textbf{\texttt{I}} &  &  $o^{2} \cdot p^{2}$   &   \\
         &         & ~\textbf{\texttt{XX}}  & ~$\Rightarrow$  & \textbf{\texttt{I}} & 0\texttt{x}0 & $\times$ & CC  \textit{vs.}  R$_{\mathrm{S}}$  \\
         &         & ~\textbf{\texttt{II}}  &      & \textbf{\texttt{I}} &  &  $\bar{o}^{1} \cdot \bar{p}^{1}$ &    \\  \hline
         & (101) & ~\textbf{\texttt{XX}}  &      & \textbf{\texttt{I}} &  &  $o^{2} \cdot p^{2}$  &   \\
         &         & ~\textbf{\texttt{XI}}  & ~$\Rightarrow$  & \textbf{\texttt{X}} & 0\texttt{x}3 & $\times$ & F  \textit{vs.}  R$_{\mathrm{S}}$  \\
         &         & ~\textbf{\texttt{IX}}  &      & \textbf{\texttt{X}} &  &  $\bar{o}^{1} \cdot \bar{p}^{1}$ &    \\ \hline
         & (110) & ~\textbf{\texttt{XI}}  &      & \textbf{\texttt{X}} &  &  $o^{2} \cdot p^{2}$   &   \\
         &         & ~\textbf{\texttt{XX}}  & ~$\Rightarrow$  & \textbf{\texttt{I}} & 0\texttt{x}1 & $\times$ & F  \textit{vs.}  R$_{\mathrm{S}}$  \\
         &         & ~\textbf{\texttt{IX}}  &      & \textbf{\texttt{X}} &  &  $\bar{o}^{1} \cdot \bar{p}^{1}$ &    \\ \hline
         & (111) & ~\textbf{\texttt{XX}}  &      & \textbf{\texttt{I}} &  &  $o^{2} \cdot p^{3}$  &   \\
         &         & ~\textbf{\texttt{XX}}  & ~$\Rightarrow$  & \textbf{\texttt{I}} & 0\texttt{x}2 & $\times$ & CC  \textit{vs.}  R$_{\mathrm{S}}$  \\
         &         & ~\textbf{\texttt{IX}}  &      & \textbf{\texttt{X}} &  &  $\bar{o}^{1} \cdot \bar{p}^{0}$ &    \\ \hline
    \end{tabular}
    %\caption{Table of.}
    %\label{tab:all_64_cases}
%\end{table}
%\phantom{-|-}
%\begin{table}[h]
    \small
    \centering
    \begin{tabular}{|cc|ccc|c|c|c|}
    \hline
\multicolumn{2}{|c|}{\textbf{Errors}} & \multicolumn{3}{c|}{\textbf{Gates}} & \textbf{Synd.} & \textbf{Prob.} & \textbf{Outcome} \\
\multicolumn{2}{|c|}{[Enviro.]} & \multicolumn{3}{c|}{Col. 5~\&~6} & Ancilla & Error $\times$ & Standard \\
\multicolumn{2}{|c|}{(Entangle)} & \multicolumn{3}{c|}{$\Rightarrow$ Result} & c-reg. & Non-error &  \textit{vs.} Assisted  \\
    \hline \hline
~[101]  & (000) & ~\textbf{\texttt{XI}}  &      & \textbf{\texttt{X}} &  &  $o^{2} \cdot p^{0}$  &  \phantom{CC  \textit{vs.}  R$_{\mathrm{PT}}$} \\
         &         & ~\textbf{\texttt{II}}  & ~$\Rightarrow$  & \textbf{\texttt{I}} & 0\texttt{x}1 & $\times$ & F  \textit{vs.}  R$_{\mathrm{S}}$  \\
         &         & ~\textbf{\texttt{XI}}  &      & \textbf{\texttt{X}} &  &  $\bar{o}^{1} \cdot \bar{p}^{3}$ &    \\ \hline
         & (001) & ~\textbf{\texttt{XX}}  &      & \textbf{\texttt{I}} &  &  $o^{2} \cdot p^{1}$   &   \\
         &         & ~\textbf{\texttt{II}}  & ~$\Rightarrow$  & \textbf{\texttt{I}} & 0\texttt{x}2 & $\times$ & CC  \textit{vs.}  R$_{\mathrm{S}}$  \\
         &         & ~\textbf{\texttt{XI}}  &      & \textbf{\texttt{X}} &  &  $\bar{o}^{1} \cdot \bar{p}^{2}$ &    \\ \hline
         & (010) & ~\textbf{\texttt{XI}}  &      & \textbf{\texttt{X}} &  &  $o^{2} \cdot p^{1}$   &   \\
         &         & ~\textbf{\texttt{IX}}  & ~$\Rightarrow$  & \textbf{\texttt{X}} & 0\texttt{x}0 & $\times$ & F  \textit{vs.}  R$_{\mathrm{S}}$  \\
         &         & ~\textbf{\texttt{XI}}  &      & \textbf{\texttt{X}} &  &  $\bar{o}^{1} \cdot \bar{p}^{2}$ &    \\ \hline
         & (100) & ~\textbf{\texttt{XI}}  &      & \textbf{\texttt{X}} &  &  $o^{2} \cdot p^{1}$   &   \\
         &         & ~\textbf{\texttt{II}}  & ~$\Rightarrow$  & \textbf{\texttt{I}} & 0\texttt{x}3 & $\times$ & CC  \textit{vs.}  R$_{\mathrm{S}}$  \\
         &         & ~\textbf{\texttt{XX}}  &      & \textbf{\texttt{I}} &  &  $\bar{o}^{1} \cdot \bar{p}^{2}$ &    \\ \hline
         & (011) & ~\textbf{\texttt{XX}}  &      & \textbf{\texttt{I}} &  &  $o^{2} \cdot p^{2}$   &   \\
         &         & ~\textbf{\texttt{IX}}  & ~$\Rightarrow$  & \textbf{\texttt{X}} & 0\texttt{x}3 & $\times$ & F  \textit{vs.}  R$_{\mathrm{S}}$  \\
         &         & ~\textbf{\texttt{XI}}  &      & \textbf{\texttt{X}} &  &  $\bar{o}^{1} \cdot \bar{p}^{1}$ &    \\ \hline
         & (101) & ~\textbf{\texttt{XX}}  &      & \textbf{\texttt{I}} &  &  $o^{2} \cdot p^{2}$  &   \\
         &         & ~\textbf{\texttt{II}}  & ~$\Rightarrow$  & \textbf{\texttt{I}} & 0\texttt{x}0 & $\times$ & CC  \textit{vs.}  R$_{\mathrm{S}}$  \\
         &         & ~\textbf{\texttt{XX}}  &      & \textbf{\texttt{I}} &  &  $\bar{o}^{1} \cdot \bar{p}^{1}$ &    \\ \hline
         & (110) & ~\textbf{\texttt{XI}}  &      & \textbf{\texttt{X}} &  &  $o^{2} \cdot p^{2}$  &   \\
         &         & ~\textbf{\texttt{IX}}  & ~$\Rightarrow$  & \textbf{\texttt{X}} & 0\texttt{x}2 & $\times$ & F  \textit{vs.}  R$_{\mathrm{S}}$  \\
         &         & ~\textbf{\texttt{XX}}  &      & \textbf{\texttt{I}} &  &  $\bar{o}^{1} \cdot \bar{p}^{1}$ &    \\ \hline
         & (111) & ~\textbf{\texttt{XX}}  &      & \textbf{\texttt{I}} &  &  $o^{2} \cdot p^{3}$  &   \\
         &         & ~\textbf{\texttt{IX}}  & ~$\Rightarrow$  & \textbf{\texttt{X}} & 0\texttt{x}1 & $\times$ & CC  \textit{vs.}  R$_{\mathrm{S}}$  \\
         &         & ~\textbf{\texttt{XX}}  &      & \textbf{\texttt{I}} &  &  $\bar{o}^{1} \cdot \bar{p}^{0}$ &    \\ \hline
    \end{tabular}
    %\caption{Table of.}
    %\label{tab:all_64_cases}
%\end{table}
%\phantom{-|-}
%\begin{table}[h]
    \small
    \centering
    \begin{tabular}{|cc|ccc|c|c|c|}
    \hline
\multicolumn{2}{|c|}{\textbf{Errors}} & \multicolumn{3}{c|}{\textbf{Gates}} & \textbf{Synd.} & \textbf{Prob.} & \textbf{Outcome} \\
\multicolumn{2}{|c|}{[Enviro.]} & \multicolumn{3}{c|}{Col. 5~\&~6} & Ancilla & Error $\times$ & Standard \\
\multicolumn{2}{|c|}{(Entangle)} & \multicolumn{3}{c|}{$\Rightarrow$ Result} & c-reg. & Non-error &  \textit{vs.} Assisted  \\
    \hline \hline
~[110]  & (000) & ~\textbf{\texttt{II}}  &      & \textbf{\texttt{I}} &  &  $o^{2} \cdot p^{0}$  &  \phantom{CC  \textit{vs.}  R$_{\mathrm{PT}}$} \\
         &         & ~\textbf{\texttt{XI}}  & ~$\Rightarrow$  & \textbf{\texttt{X}} & 0\texttt{x}3 & $\times$ & F  \textit{vs.}  R$_{\mathrm{S}}$  \\
         &         & ~\textbf{\texttt{XI}}  &      & \textbf{\texttt{X}} &  &  $\bar{o}^{1} \cdot \bar{p}^{3}$ &    \\ \hline
         & (001) & ~\textbf{\texttt{IX}}  &      & \textbf{\texttt{X}} &  &  $o^{2} \cdot p^{1}$  &   \\
         &         & ~\textbf{\texttt{XI}}  & ~$\Rightarrow$  & \textbf{\texttt{X}} & 0\texttt{x}0 & $\times$ & F  \textit{vs.}  R$_{\mathrm{S}}$  \\
         &         & ~\textbf{\texttt{XI}}  &      & \textbf{\texttt{X}} &  &  $\bar{o}^{1} \cdot \bar{p}^{2}$ &    \\ \hline
         & (010) & ~\textbf{\texttt{II}}  &      & \textbf{\texttt{I}} &  &  $o^{2} \cdot p^{1}$   &   \\
         &         & ~\textbf{\texttt{XX}}  & ~$\Rightarrow$  & \textbf{\texttt{I}} & 0\texttt{x}2 & $\times$ & CC  \textit{vs.}  R$_{\mathrm{S}}$  \\
         &         & ~\textbf{\texttt{XI}}  &      & \textbf{\texttt{X}} &  &  $\bar{o}^{1} \cdot \bar{p}^{2}$ &    \\ \hline
         & (100) & ~\textbf{\texttt{II}}  &      & \textbf{\texttt{I}} &  &  $o^{2} \cdot p^{1}$  &   \\
         &         & ~\textbf{\texttt{XI}}  & ~$\Rightarrow$  & \textbf{\texttt{X}} & 0\texttt{x}1 & $\times$ & CC  \textit{vs.}  R$_{\mathrm{S}}$  \\
         &         & ~\textbf{\texttt{XX}}  &      & \textbf{\texttt{I}} &  &  $\bar{o}^{1} \cdot \bar{p}^{2}$ &    \\ \hline
         & (011) & ~\textbf{\texttt{IX}}  &      & \textbf{\texttt{X}} &  &  $o^{2} \cdot p^{2}$  &   \\
         &         & ~\textbf{\texttt{XX}}  & ~$\Rightarrow$  & \textbf{\texttt{I}} & 0\texttt{x}1 & $\times$ & F  \textit{vs.}  R$_{\mathrm{S}}$  \\
         &         & ~\textbf{\texttt{XI}}  &      & \textbf{\texttt{X}} &  &  $\bar{o}^{1} \cdot \bar{p}^{1}$ &    \\ \hline
         & (101) & ~\textbf{\texttt{IX}}  &      & \textbf{\texttt{X}} &  &  $o^{2} \cdot p^{2}$  &   \\
         &         & ~\textbf{\texttt{XI}}  & ~$\Rightarrow$  & \textbf{\texttt{X}} & 0\texttt{x}2 & $\times$ & F  \textit{vs.}  R$_{\mathrm{S}}$  \\
         &         & ~\textbf{\texttt{XX}}  &      & \textbf{\texttt{I}} &  &  $\bar{o}^{1} \cdot \bar{p}^{1}$ &    \\ \hline
         & (110) & ~\textbf{\texttt{II}}  &      & \textbf{\texttt{I}} &  &  $o^{2} \cdot p^{2}$   &   \\
         &         & ~\textbf{\texttt{XX}}  & ~$\Rightarrow$  & \textbf{\texttt{I}} & 0\texttt{x}0 & $\times$ & CC  \textit{vs.}  R$_{\mathrm{S}}$  \\
         &         & ~\textbf{\texttt{XX}}  &      & \textbf{\texttt{I}} &  &  $\bar{o}^{1} \cdot \bar{p}^{1}$ &    \\ \hline
         & (111) & ~\textbf{\texttt{IX}}  &      & \textbf{\texttt{X}} &  &  $o^{2} \cdot p^{3}$   &   \\
         &         & ~\textbf{\texttt{XX}}  & ~$\Rightarrow$  & \textbf{\texttt{I}} & 0\texttt{x}3 & $\times$ & CC  \textit{vs.}  R$_{\mathrm{S}}$  \\
         &         & ~\textbf{\texttt{XX}}  &      & \textbf{\texttt{I}} &  &  $\bar{o}^{1} \cdot \bar{p}^{0}$ &    \\ \hline
    \end{tabular}
    %\caption{Table of.}
    %\label{tab:all_64_cases}
%\end{table}
%\phantom{-|-}
%\begin{table}[h]
    \small
    \centering
    \begin{tabular}{|cc|ccc|c|c|c|}
    \hline
\multicolumn{2}{|c|}{\textbf{Errors}} & \multicolumn{3}{c|}{\textbf{Gates}} & \textbf{Synd.} & \textbf{Prob.} & \textbf{Outcome} \\
\multicolumn{2}{|c|}{[Enviro.]} & \multicolumn{3}{c|}{Col. 5~\&~6} & Ancilla & Error $\times$ & Standard \\
\multicolumn{2}{|c|}{(Entangle)} & \multicolumn{3}{c|}{$\Rightarrow$ Result} & c-reg. & Non-error &  \textit{vs.} Assisted  \\
    \hline \hline
~[111]  & (000) & ~\textbf{\texttt{XI}}  &      & \textbf{\texttt{X}} &  &  $o^{3} \cdot p^{0}$  &  \phantom{CC  \textit{vs.}  R$_{\mathrm{PT}}$} \\
         &         & ~\textbf{\texttt{XI}}  & ~$\Rightarrow$  & \textbf{\texttt{X}} & 0\texttt{x}0 & $\times$ & F  \textit{vs.}  R$_{\mathrm{S}}$  \\
         &         & ~\textbf{\texttt{XI}}  &      & \textbf{\texttt{X}} &  &  $\bar{o}^{0} \cdot \bar{p}^{3}$ &    \\ \hline
         & (001) & ~\textbf{\texttt{XX}}  &      & \textbf{\texttt{I}} &  &  $o^{3} \cdot p^{1}$  &   \\
         &         & ~\textbf{\texttt{XI}}  & ~$\Rightarrow$  & \textbf{\texttt{X}} & 0\texttt{x}3 & $\times$ & F  \textit{vs.}  R$_{\mathrm{S}}$  \\
         &         & ~\textbf{\texttt{XI}}  &      & \textbf{\texttt{X}} &  &  $\bar{o}^{0} \cdot \bar{p}^{2}$ &    \\ \hline
         & (010) & ~\textbf{\texttt{XI}}  &      & \textbf{\texttt{X}} &  &  $o^{3} \cdot p^{1}$   &   \\
         &         & ~\textbf{\texttt{XX}}  & ~$\Rightarrow$  & \textbf{\texttt{I}} & 0\texttt{x}1 & $\times$ & F  \textit{vs.}  R$_{\mathrm{S}}$  \\
         &         & ~\textbf{\texttt{XI}}  &      & \textbf{\texttt{X}} &  &  $\bar{o}^{0} \cdot \bar{p}^{2}$ &    \\ \hline
         & (100) & ~\textbf{\texttt{XI}}  &      & \textbf{\texttt{X}} &  &  $o^{3} \cdot p^{1}$   &   \\
         &         & ~\textbf{\texttt{XI}}  & ~$\Rightarrow$  & \textbf{\texttt{X}} & 0\texttt{x}2 & $\times$ & F  \textit{vs.}  R$_{\mathrm{S}}$  \\
         &         & ~\textbf{\texttt{XX}}  &      & \textbf{\texttt{I}} &  &  $\bar{o}^{0} \cdot \bar{p}^{2}$ &    \\ \hline
         & (011) & ~\textbf{\texttt{XX}}  &      & \textbf{\texttt{I}} &  &  $o^{3} \cdot p^{2}$  &   \\
         &         & ~\textbf{\texttt{XX}}  & ~$\Rightarrow$  & \textbf{\texttt{I}} & 0\texttt{x}2 & $\times$ & CC  \textit{vs.}  R$_{\mathrm{S}}$  \\
         &         & ~\textbf{\texttt{XI}}  &      & \textbf{\texttt{X}} &  &  $\bar{o}^{0} \cdot \bar{p}^{1}$ &    \\ \hline
         & (101) & ~\textbf{\texttt{XX}}  &      & \textbf{\texttt{I}} &  &  $o^{3} \cdot p^{2}$   &   \\
         &         & ~\textbf{\texttt{XI}}  & ~$\Rightarrow$  & \textbf{\texttt{X}} & 0\texttt{x}1 & $\times$ & CC  \textit{vs.}  R$_{\mathrm{S}}$  \\
         &         & ~\textbf{\texttt{XX}}  &      & \textbf{\texttt{I}} &  &  $\bar{o}^{0} \cdot \bar{p}^{1}$ &    \\ \hline
         & (110) & ~\textbf{\texttt{XI}}  &      & \textbf{\texttt{X}} &  &  $o^{3} \cdot p^{2}$  &   \\
         &         & ~\textbf{\texttt{XX}}  & ~$\Rightarrow$  & \textbf{\texttt{I}} & 0\texttt{x}3 & $\times$ & CC  \textit{vs.}  R$_{\mathrm{S}}$  \\
         &         & ~\textbf{\texttt{XX}}  &      & \textbf{\texttt{I}} &  &  $\bar{o}^{0} \cdot \bar{p}^{1}$ &    \\ \hline
         & (111) & ~\textbf{\texttt{XX}}  &      & \textbf{\texttt{I}} &  &  $o^{3} \cdot p^{3}$  &   \\
         &         & ~\textbf{\texttt{XX}}  & ~$\Rightarrow$  & \textbf{\texttt{I}} & 0\texttt{x}0 & $\times$ & CC  \textit{vs.}  R$_{\mathrm{S}}$  \\
         &         & ~\textbf{\texttt{XX}}  &      & \textbf{\texttt{I}} &  &  $\bar{o}^{0} \cdot \bar{p}^{0}$ &    \\ \hline
    \end{tabular}
    \caption{See main report and Table~\ref{tab:001_cases} for description of tables. Here R$_{\mathrm{S}}$ = REJECT based on sensors.}
    \label{tab:all_64_cases_II}
\end{table*}

\end{document}